\documentclass[iop]{emulateapj}
\RequirePackage{epstopdf}
\usepackage{multirow}
\RequirePackage{graphicx}
\usepackage{lineno}

\newcommand{\kepler}{{\it Kepler }}

\newcommand{\um}{$\mu$m}

\newcommand{\fbol}{$F_{\mathrm{BOL}}$}
\newcommand{\teff}{$T_{\mathrm{eff}}$}
\newcommand{\teffbol}{$T_{\mathrm{BOL}}$}
\newcommand{\teffbt}{$T_{\mathrm{PH}}$}

\slugcomment{ApJ in review, revised \today}

\shorttitle{Temperatures of M dwarfs}
\shortauthors{Mann, Gaidos, \& Ansdell}

\begin{document}

\title{Spectro-Thermometry of M dwarfs and their candidate planets:\\ too hot, too cool, or just right?}

\author{Andrew W. Mann\altaffilmark{1,2}, Eric Gaidos\altaffilmark{3}, Megan Ansdell\altaffilmark{1}}
  
\altaffiltext{1}{Institute for Astronomy, University of Hawai'i, 2680 Woodlawn Drive, Honolulu, HI 96822} 
\altaffiltext{2}{Harlan J. Smith Fellow, Department of Astronomy, The University of Texas at Austin, Austin, TX 78712, USA} 
\altaffiltext{3}{Department of Geology \& Geophysics, University of Hawai'i, 1680 East-West Road, Honolulu, HI 96822} 

\begin{abstract}
We use moderate-resolution spectra of nearby late K and M dwarf stars with parallaxes and interferometrically determined radii to refine their effective temperatures, luminosities, and metallicities. We use these revised values to calibrate spectroscopic techniques to infer the fundamental parameters of more distant late-type dwarf stars. We demonstrate that, after masking out poorly modeled regions, the newest version of the PHOENIX atmosphere models accurately reproduce temperatures derived bolometrically. We apply methods to late-type hosts of transiting planet candidates in the \kepler field, and calculate effective temperature, radius, mass, and luminosity with typical errors of 57~K, 7\%, 11\%, and 13\%, respectively. We find systematic offsets between our values and those from previous analyses of the same stars, which we attribute to differences in atmospheric models utilized for each study. We investigate which of the planets in this sample are likely to orbit in the circumstellar habitable zone. We determine that four candidate planets (KOI 854.01, 1298.02, 1686.01, and 2992.01) are inside of or within 1$\sigma$ of a conservative definition of the habitable zone, but that several planets identified by previous analyses are not (e.g. KOI 1422.02 and KOI 2626.01). Only one of the four habitable-zone planets is Earth sized, suggesting a downward revision in the occurrence of such planets around M dwarfs. These findings highlight the importance of measuring accurate stellar parameters when deriving parameters of their orbiting planets.
\end{abstract}

\keywords{stars: fundamental parameters --- stars: late-type --- planetary systems --- astrobiology}

\section{Introduction}\label{sec:intro} 
M dwarf stars have become prime targets in the search for potentially habitable planets in large part because they are much cooler and smaller than Sun-like stars, allowing smaller planets in the habitable zone (HZ) to be detected. Observations have shown that the occurrence of small planets -- including Earth- and super Earth-sized planets -- increases with decreasing stellar mass \citep{Howard:2012yq} \citep[although also see][]{Fressin:2013qy}, and that small planets are nearly ubiquitously around late-type stars \citep{Swift:2013vn}. Since over 70\% of stars in the solar neighborhood are M dwarfs \citep{Henry:1994fk, Chabrier:2003fk, Reid:2004lr} it is likely that the vast majority of potentially habitable planets near our Sun orbit around late K and M dwarfs.

For a fixed planet size and radius, the Doppler signal is inversely related to stellar mass, $M_*$, and the planet's orbital period, $P$ ($\propto M_*^{-2/3}P^{-1/3}$), while the transit depth is inversely related to the stellar radius, $R_*$ ($\propto R_*^{-2}$). The transit probability scales inversely to the planet's orbital period ($\propto R_*M_*^{-1/3}P^{-2/3}$). An early M dwarf has about half the radius and mass of a Sun-like star. A planet in the HZ (the range of stellar irradiances bracketed by the runaway greenhouse on one end and the first CO$_2$ condensation at the other) of an early M dwarf has $\sim 1/6$th the period of a HZ planet around a Sun-like star \citep{2007Icar..191..453S, Kopparapu2013}. Thus a planet in the HZ of an early M dwarf will have $\sim$3 times the Doppler signal, $\sim$4 times the transit depth, and $\sim$2 times the probability of transiting compared to a HZ planet around a solar-type star. 

The \kepler mission \citep{Borucki:2010lr} has found more than 2000 planet candidates \citep{Borucki:2011uq,Batalha:2013lr}, enabling more robust studies of exoplanet statistics through large samples. Although \kepler target stars are mostly F, G, and early K dwarfs \citep{Batalha:2010fk}, there are also $\sim$4000 \kepler stars cooler than 4000~K \citep{Dressing:2013fk}, which collectively harbor more than 100 detected candidate exoplanets. This is roughly an order of magnitude more planets than have been found around M dwarfs using ground-based Doppler \citep{Bonfils:2013ys} or transit \citep{Berta:2013aa} surveys. As a result, \kepler has been critical in constraining the occurrence of planets around M dwarfs \citep{Mann:2012,Dressing:2013fk,Swift:2013vn}, the habitability of planets orbiting M dwarfs \citep{Dressing:2013fk,Kopparapu2013b,Gaidos:2013rt}, the metallicity distribution of M dwarfs with or without detected planets \citep{Muirhead:2012pd, Mann:2013vn}, and the mass and radius distributions of M dwarf planets \citep{Gaidos:2012lr,Morton:2013}.

As is the case for most \kepler discoveries, accurate physical parameters for planets are largely limited by uncertainties its host star parameters.  For example, the transit depth gives only the ratio of the planet radius to that of the star; one must know the size of the star in order to accurately determine the size of the planet. Further, whether a planet resides in the HZ depends on the irradiance ($S$) that the planet receives from its host star, which in turn depends sensitively on the host star's \teff. 

Total stellar irradiation is not the only measure of planetary habitability \citep{Gaidos2005} and planets in the HZ of M dwarfs will, at the same total stellar irradiation, experience a different environment than their counterparts orbiting solar-type stars \citep{Tarter2007}.  The spectral energy distribution of M dwarfs is distinct from that of G stars.  Because the reflectivity of water (ice) surfaces is wavelength-dependent, this alters the efficacy of the ice albedo feedback that can destabilize the climates of Earth-like planets \citep{Joshi2012,Shields2013}.  Most oxygenic photosynthesis on Earth requires light bluer than (more energetic than) $\sim$700 nm, and thus equivalent life around M dwarfs would have comparatively less light to harvest. 

Planets around late M dwarfs with persistent magnetic and chromospheric activity will experience elevated ultraviolet (UV) emission and stellar wind particle fluxes. These could heat and erode atmospheres \citep{Lammer2007} as well as drive atmospheric chemistry \citep{Segura2010}.  Although tidally induced synchronization of the rotation of planets in the HZ of M dwarfs may not lead to atmospheric collapse on the cold night side \citep{Joshi1997}, it could lead to sequestration of less volatile substances such as water ice \citep{Menou2013}.  Finally, impacts by planetesimals will be more energetic in the HZs of M dwarfs because orbital velocities are higher; this may remove volatiles such as water from the planet altogether \citep{Lissauer2007}. Although these concerns suggest caution in designating any planets in the HZ as actually ``habitable,'' they have not damped enthusiasm for identifying such planets and estimating their occurrence \citep{Dressing:2013fk,Kopparapu2013b,Gaidos:2013rt}. 

Constraining the physical parameters of M dwarfs is significantly more difficult than for their warmer FGK counterparts. Their cooler temperatures enable the appearance of molecular bands (e.g., CaH, TiO) that dominate their spectra, create line confusion, and obscure the continuum level \citep{1976A&A....48..443M, Kirkpatrick:1991kx}. The poorly constrained optical parameters and reaction constants makes these bands difficult to model, rendering spectral synthesis unreliable for M dwarfs, although improvements are ongoing \citep[e.g.,][]{Bean:2006fk, Onehag:2012lr}.

Nonetheless, the radii of bright, nearby M dwarfs can be measured using parallaxes and long-baseline interferometry \citep[e.g.,][]{Demory:2009qy, von-Braun:2011bh, von-Braun:2012lq}. \citet[][henceforth B12]{Boyajian:2012lr} use the CHARA interferometer to obtain angular diameters for a set of K and M dwarfs. They utilize {it Hipparcos} parallaxes \citep{van-Leeuwen:2005kx,van-Leeuwen:2007yq} to determine $R_*$ and combine photometry with template spectra to determine bolometric fluxes and infer \teff{} with high precision (1-2\% error in $R_*$, $< 1\%$ error in \teff). They derive $L_*$ by combining the bolometric flux with the distance (parallax), and determine $M_*$ from empirical relations between absolute $K$-band magnitude ($M_K$) and $M_*$ \citep{Henry:1993fk, 2000A&A...364..217D}. As a result, the B12 sample makes an ideal set of ``calibration'' stars to develop and calibrate techniques for measuring the physical characteristics of late K and M dwarfs. These empirical techniques can then be applied to more distant stars beyond the reach of interferometry or precision astrometry (parallaxes). 

Here we present observations and analysis of the B12 late-type stars to derive empirical relations between fundamental physical parameters and moderate-resolution spectra. We apply these relations to \kepler M dwarf planet hosts, (re)estimate planet radii and stellar irradiances, and determine which of their planets reside in the HZ. First we describe our samples of B12 calibration stars and \kepler targets in Section~\ref{sec:sample}, then we summarize our moderate-resolution visible and near infrared (NIR) spectroscopy of both samples in Section~\ref{sec:obs}. In Section~\ref{sec:fixtab} we use our spectra to re-derive bolometric fluxes, and hence physical parameters, of the B12 sample, and then derive new empirical relations between \teff{} and several other stellar parameters. In Section~\ref{sec:teff} we calibrate the \teff{} values based on comparing models to observed spectra. We apply our techniques to late-type \kepler planet candidate hosts in Section~\ref{sec:kepler}. We use these newly derived stellar parameters in Section~\ref{sec:habitable} to calculate the irradiances and radii of the candidate planets, which we utilize to investigate which \kepler exoplanets orbit in the HZ. We conclude in Section~\ref{sec:discussion} by briefly commenting on the complications and ramifications of our work.  

In this paper we will refer to \kepler exoplanet candidates as planets, both for simplicity and to avoid confusion with candidate HZ planets (planets which are potentially in the HZ). For specific systems we will refer to them using the \kepler object of interest (KOI) number, with whole numbers referring to the host star (e.g., KOI 854), and decimals to denote the planets (e.g., KOI 854.01). 

\section{Sample}\label{sec:sample}
For our calibration sample, we selected B12 stars with \teff$<4800$~K and $-45^\circ<\delta<70^\circ$ (observable with the telescopes and instruments specified in Section~\ref{sec:obs}). This cut contains 23 stars spanning $3100$~K$<$\teff$<4800$~K, which covers the expected spectral types and \teff{} of our \kepler planet host sample (defined below). 

Following the criterion of \citet{Mann:2013vn}, we selected \kepler planet hosts with $K_P-J>1.85$, excluding three suspected false positives (see below). This color cut includes all dwarfs with \teff$<4100$~K, with some as warm as $4500$~K \citep{Mann:2012}, and at least one as cool as 3200~K \citep{Muirhead:2012zr}. We included all planet candidates listed in the NASA exoplanet archive as of 2013 February \citep{Batalha:2013lr}.

In our \kepler sample, we excluded three planets that are likely to be false positives: KOI 977 is a giant star \citep{Muirhead:2012pd, Mann:2012}, KOI 1902 is an eclipsing binary \citep{Mann:2013vn}, and KOI 256 is a white dwarf-M dwarf binary \citep{Muirhead:2013lr}. The final sample includes 123 late-type dwarfs harboring 188 detected planets. 

\section{Observations and Reduction}\label{sec:obs}
Optical spectra of both the calibration sample and the \kepler planet host sample were obtained with the SuperNova Integral Field Spectrograph \citep[SNIFS,][]{2002SPIE.4836...61A,Lantz:2004} on the University of Hawaii 2.2m telescope atop Mauna Kea. SNIFS uses a dichroic mirror to separate the incoming light onto blue (3200-5200\,\AA) and red (5100-9700\,\AA) spectrograph channels. The spectral resolution, $R$, is $\simeq800$ in the blue channel and $\simeq1000$ in the red. Integration times for stars in the calibration sample varied from 5 to 40~s, which was sufficient to achieve high signal-to-noise (S/N $>200$ per resolving element) in the red channel while avoiding the non linear regime of the detector. For \kepler stars, which are fainter, total integration times were between 5 minutes and 2~hr. For targets with integration times longer than 30~m, integrations were divided into three or more exposures $\le$30 minutes long, and then stacked (using the weighted mean) to mitigate contamination from cosmic rays and other artifacts. The resulting S/N for \kepler stars is typically $>60$ (per resolving element) in the red channel. For three of the \kepler targets (KOI 2306, 1879, and 2418) we were unable to extract a reasonable spectrum from the blue channel. However, the blue channel data is not used for our analysis of the KOIs.

The SNIFS pipeline \citep{Bacon:2001,Aldering:2006} performed dark, bias, and flat-field corrections and cleaned the data of bad pixels and cosmic rays. The pipeline wavelength calibrated the data based on arc lamp exposures taken at the same telescope pointing and time as the science data. Over the course of each night, we obtained spectra of the EG131, Feige 66, Feige 110, BD+284211, or BD+174708 spectrophotometric standards \citep{Bessell:1999fk,Bohlin:2001fj,Hamuy:1992qy,Oke:1990}. We combined the standard star observations with a model of the atmosphere above Mauna Kea from \citet{Buton:2013lr} to correct each spectrum for instrument response and atmospheric extinction and to remove telluric lines. We shifted each spectrum in wavelength to the rest frames of their source stars by cross-correlating each spectrum to a spectral template of similar spectral type from the Sloan Digital Sky Survey \citep{Stoughton:2002, Bochanski:2007lr} using the IDL routine {\it xcorl} \citep{Mohanty:2003qy, West:2009fk}. More details on our data reduction can be found in \citet{Mann:2012} and \citet{Lepine:2013lr}. 

Multiple observations of the same spectrophotometric standards demonstrate that SNIFS spectra have an error term of $\simeq1\%$ in addition to the expected (mostly Poisson) measurement noise \citep{Mann:2011qy,Buton:2013lr}. For our calibrator stars this source of error is larger than that from the Poisson noise, and is therefore included in our analysis.

We obtained NIR spectra of our B12 calibration sample and selected \kepler stars with candidate HZ planets (see Section~\ref{sec:habitable}) using the SpeX spectrograph \citep{Rayner:2003lr} attached to the NASA Infrared Telescope Facility (IRTF) on Mauna Kea. SpeX observations were taken in the short cross-dispersed (SXD) mode using the 0.3$\times15\arcsec$ slit, yielding simultaneous coverage from 0.8 to 2.4\um\ at a resolution of $R\simeq2000$. A target was placed at two positions along the slit (A and B) and observed in an ABBA pattern to accurately subtract the sky background by differencing. For each B12 star we took 6 exposures in this pattern, which gave a S/N $>150$ in the $H$- and $K$-bands for all targets (and typically $>200$). For the \kepler HZ candidate hosts a total of 10-30 exposures were taken to achieve a peak S/N $\gtrsim 60$ in the $H$- and $K$-bands. To remove effects from large telescope slews, we obtained flat-field and argon lamp calibration sequences after each target. 

Spectra were extracted using the SpeXTool package \citep{Cushing:2004fk}, which performed flat-field correction, wavelength calibration, sky subtraction, and extraction of the one-dimensional spectrum. Multiple exposures were combined using the IDL routine {\it xcombxpec}. To correct for telluric lines, we observed an A0V-type star within 1~hr and 0.1 airmass of the target observation (usually much closer in time and airmass). In the cases where a target's airmass changed by $\ge 0.2$ or a single star observation lasted $>1$~hr two A0V stars were obtained. A telluric correction spectrum was constructed from each A0V star and applied to the relevant spectrum using the {\it xtellcor} package \citep{Vacca:2003qy}. We placed each spectrum in its star's rest frame by cross-correlating it with a spectrum of the template M1.5 dwarf HD 36395 from the IRTF spectral library \citep{Cushing:2005lr, Rayner:2009kx}. 

\citet{Rayner:2009kx} find that, when using the 0.3\arcsec slit in SXD mode, time-dependent changes in seeing, guiding, and differential atmospheric refraction can cause changes in the slope of SpeX data by as much as $\simeq2\%$. We mitigated this effect by observing at the parallactic angle and minimizing the time between target and standard star observations. We used the change in slope between unstacked observations of the same star as a measure of the size of this error term for each target, which for our stars is typically $<1\%$. This change in slope is significant for observations of our calibration sample, and thus is included in our error analysis.

Wavelengths in this paper are all reported in vacuum. 

\section{Modifications to the Stellar Parameters of B12}\label{sec:fixtab}
\subsection{Bolometric Flux, Effective Temperature, and Luminosity}\label{sec:reviseteff}
B12 measure the \teff{} of their sample following the method outlined in \citet{van-Belle:2008lr}. They use $BVRIJHK$ photometry to find the best-fit template spectra (or combination of template spectra) from the \citet{1998PASP..110..863P} library as well as the normalization constants for the spectra. Spectra for most of the late K and M dwarfs from \citet{1998PASP..110..863P} extend to $\simeq1.1$\um, beyond which \citet{van-Belle:2008lr} and B12 extrapolate using NIR (usually $JHK$) photometry. They calculate \fbol{} values from the normalized spectrum using the formula:
\begin{equation}\label{eqn:fbol}
\mathrm{F_{BOL}} =  \int_0^\infty F_\lambda d\lambda,
\end{equation}
where $F_\lambda$ is the flux density at wavelength $\lambda$. From there B12 calculate \teff{} of a given star using the Stefan--Boltzmann law:
\begin{equation}\label{eqn:teff_fbol}
T_\mathrm{{eff}} = 2341\left(\frac{\mathrm{F_{BOL}}}{\theta_{\mathrm{LD}}^2}\right)^{\frac{1}{4}},
\end{equation}
where \teff{} is given in Kelvin, \fbol{} is in units of $10^{-8}$ erg cm$^{-2}$ s$^{-1}$, and $\theta_{\mathrm{LD}}$ is the measured angular diameter of the star corrected for limb darkening and in units of milli-arcseconds. 

It is straightforward to calculate the stellar luminosity ($L_*$) for stars with a known distance (parallax) and \fbol{} according to the formula:
\begin{equation}\label{eqn:lum}
L = 4\pi d^2 \times\mathrm{F_{BOL}},
\end{equation}
where $d$ is the distance, which B12 draw from {\it Hipparcos} parallaxes \citep{van-Leeuwen:2005kx,van-Leeuwen:2007yq}. For all above calculations B12 assume no extinction along the line of site, which is reasonable given that these stars are nearby. Thus we make the same assumption.

Rather than employ template spectra, we calculate \fbol{} from the actual spectra we obtained of each star. Although our combined SNIFS and SpeX spectra cover the wavelengths where K and M dwarfs emit the majority of their flux, SpeX spectra have a small gap at $\simeq 1.8$\um{} and several of our stars have comparatively low S/N ($<30$) in regions of telluric contamination. We replace these regions using the best-fit PHOENIX BT-SETTL model atmosphere \citep{Allard:2011lr} following the procedure described in \citet{Mann:2013vn} and \citet{Lepine:2013lr}, and summarized in Section~\ref{sec:modelfit}. The resulting spectra cover 0.2--3.0~\um. Blueward of this region we assume the flux follows Wein's approximation, and redward we assume it follows the Rayleigh--Jeans law, and which we fit for using the 0.2--0.4\um{} and 2.0--3.0\um\ regions, respectively. Note that our typical stars have very little flux outside of 0.3--2.4\um, thus our approximation of the spectrum in this region has a negligible effect on our results.

Variations in atmospheric transparency and observing conditions between target and standard star observations as well as our choice of spectrophotometric standards \citep[e.g., the EG131 standard is slightly variable,][]{Wickramasinghe:1978} can change the overall flux level (the absolute flux calibration) of each spectrum even while preserving the relative flux calibration. An erroneous overall flux level will lead directly to an erroneous \fbol{} value, so we correct for this using visible and infrared photometry from the literature \citep[e.g.,][]{Mermilliod:1997qe}. Sources of photometry from each star are listed in Table~\ref{tab:sample}. Following the technique of \citet{Rayner:2009kx}, for each photometric point we calculate the ratio of the photometric flux to the equivalent flux synthesized from the spectra, $C_i$;
\begin{equation}
C_i = \frac{f_{\mathrm{zp},i}\times10^{-0.4m_i}}{(\int_\lambda F_\lambda \times S_{\lambda,i} d\lambda)},
\end{equation}
where $f_{\mathrm{zp}}$ is the band zero point, $m_i$ is the apparent magnitude, and $S_{\lambda,i}$ is the filter transmission for a given photometric band, $i$. Errors in $C_i$ account for (uncorrelated) errors in $m_i$ and $F_\lambda$, but $f_{\mathrm{zp},i}$ and $S_{\lambda,i}$ are assumed to have negligible uncertainty. The final correction factor ($C_*$) applied to each star's spectrum is the error-weighted mean of the $C_i$ values for a given star, which has error: 
\begin{equation}\label{eqn:cerr}
\sigma_{C_*}^2 = \frac{1}{\sum_{i}\sigma_{C_i}^2}.
\end{equation}
If errors in the photometry or spectroscopy are underestimated, or if different photometric sources are systematically offset from each other, Equation~\ref{eqn:cerr} will be an underestimate of the total error in $C_*$. For this reason we also calculate the reduced $\chi^2$ ($\chi_{\mathrm{red}}^2$) values for our correction factors (listed in Table~\ref{tab:sample}). For all but one star (GJ 725B) we get $\chi_{\mathrm{red}}^2 <3$ (and most are $<2$). This indicates that our derived errors properly account for the scatter in $C_i$ values for all targets except GJ 725B, which has a $\chi_{\mathrm{red}}^2$ of 7.4.

\begin{deluxetable*}{l l l l l l l l l l l l l l l}
\tablecaption{Calibration Sample}
\tablewidth{0pt}
\tablehead{
\colhead{Name} & \colhead{\fbol\ (10$^{-8}$)} & \colhead{$\sigma_{\mathrm{F_{BOL}}}$ } & \colhead{\teff} & \colhead{$\sigma_{T_{\mathrm{eff}}}$} & \colhead{$R_*$} & \colhead{$\sigma_{R_*}$} & \colhead{$M_*$} & \colhead{$\sigma_{M_*}$} & \colhead{$L_*$} & \colhead{$\sigma_{L_*}$} & \colhead{[Fe/H]\tablenotemark{a}} & \colhead{$\chi^2_\mathrm{{red}}$\tablenotemark{b}} & \colhead{Phot Ref\tablenotemark{c}}  \\
\colhead{} & \colhead{\scriptsize erg cm$^{-2}$ s$^{-1}$} & \colhead{} & \colhead{K} & \colhead{}  & \colhead{$R_\odot$} & \colhead{} & \colhead{$M_\odot$} & \colhead{} & \colhead{$L_\odot$} & \colhead{} & \colhead{} & \colhead{} & \colhead{}  
}
\startdata
GJ 15A &  5.664 & 0.067 & 3602 & 13 & 0.3863 & 0.0021 & 0.405 & 0.041 & 0.02256 & 0.00027 & $-0.30$ &  2.2 & 1,2,4,5 \\
GJ 205 &  6.510 & 0.133 & 3850 & 22 & 0.5735 & 0.0044 & 0.637 & 0.064 & 0.06449 & 0.00139 & $+0.49$ &  2.8 & 1,2,6,3,4,7 \\
GJ 380 & 15.200 & 0.224 & 4176 & 19 & 0.6398 & 0.0046 & 0.711 & 0.071 & 0.11174 & 0.00167 & $+0.24$ &  0.9 & 1,2,4,5 \\
GJ 526 &  4.105 & 0.052 & 3646 & 34 & 0.4840 & 0.0084 & 0.490 & 0.049 & 0.03694 & 0.00051 & $-0.31$ &  1.6 & 1,2,6,3,4,5 \\
GJ 687 &  3.511 & 0.067 & 3457 & 35 & 0.4183 & 0.0070 & 0.403 & 0.040 & 0.02228 & 0.00044 & $-0.05$ &  1.8 & 2,8,9 \\
GJ 880 &  3.572 & 0.032 & 3731 & 16 & 0.5477 & 0.0048 & 0.572 & 0.057 & 0.05181 & 0.00058 & $+0.21$ &  2.3 & 1,2,6,3,10,11 \\
GJ 887 & 11.118 & 0.225 & 3695 & 35 & 0.4712 & 0.0086 & 0.494 & 0.049 & 0.03694 & 0.00075 & $-0.06$ &  0.9 & 1,2,3,12,7 \\
GJ 699 &  3.319 & 0.032 & 3238 & 11 & 0.1869 & 0.0012 & 0.159 & 0.016 & 0.00342 & 0.00003 & $-0.40$ &  1.8 & 1,2,6,4,7 \\
GJ 411 & 10.627 & 0.146 & 3532 & 17 & 0.3924 & 0.0033 & 0.392 & 0.039 & 0.02134 & 0.00030 & $-0.38$ &  2.0 & 1,2,3,13,5 \\
GJ 105A & 17.302 & 0.216 & 4704 & 21 & 0.7949 & 0.0062 & 0.767 & 0.124 & 0.27615 & 0.00356 & $-0.28$ &  1.2 & 1,2,3,14,5 \\
GJ 338A &  6.171 & 0.119 & 3953 & 41 & 0.5773 & 0.0131 & 0.630 & 0.063 & 0.07316 & 0.00276 & $-0.01$ &  2.7 & 2,15,16 \\
GJ 338B &  5.799 & 0.044 & 3926 & 37 & 0.5673 & 0.0137 & 0.617 & 0.062 & 0.06875 & 0.00229 & $-0.04$ &  2.2 & 1,15,15,16 \\
GJ 412A &  3.044 & 0.029 & 3537 & 41 & 0.3982 & 0.0091 & 0.390 & 0.039 & 0.02214 & 0.00024 & $-0.37$ &  2.9 & 1,2,4,5 \\
GJ 436 &  0.874 & 0.009 & 3520 & 66 & 0.4546 & 0.0182 & 0.447 & 0.045 & 0.02834 & 0.00072 & $+0.01$ &  1.5 & 1,17,18 \\
GJ 570A & 20.463 & 0.278 & 4588 & 58 & 0.7390 & 0.0190 & 0.740 & 0.119 & 0.21609 & 0.00317 & $-0.06$ &  2.2 & 1,2,4,7 \\
GJ 581 &  0.971 & 0.011 & 3487 & 62 & 0.2990 & 0.0100 & 0.308 & 0.031 & 0.01181 & 0.00021 & $-0.15$ &  1.1 & 1,2,6,4,7 \\
GJ 702B & 19.922 & 0.474 & 4475 & 33 & 0.6697 & 0.0089 & 0.749 & 0.075 & 0.15953 & 0.00396 & $+0.01$ &  2.5 & 2,14,19 \\
GJ 725A &  3.989 & 0.038 & 3417 & 17 & 0.3561 & 0.0039 & 0.330 & 0.033 & 0.01573 & 0.00019 & $-0.23$ &  2.9 & 1,2,20,8,16 \\
GJ 725B\tablenotemark{d} &  2.351 & 0.023 & 3142 & 29 & 0.3232 & 0.0061 & 0.257 & 0.026 & 0.00927 & 0.00011 & $-0.30$ &  7.4 & 1,2,3,8,16 \\
GJ 809 &  3.413 & 0.042 & 3744 & 27 & 0.5472 & 0.0067 & 0.573 & 0.057 & 0.05250 & 0.00069 & $-0.06$ &  2.3 & 1,2,10,9 \\
GJ 820A & 38.436 & 0.506 & 4399 & 16 & 0.6611 & 0.0048 & 0.727 & 0.073 & 0.14606 & 0.00196 & $-0.27$ &  1.3 & 1,2,14,11 \\
GJ 820B & 22.284 & 0.323 & 4025 & 24 & 0.6010 & 0.0072 & 0.656 & 0.066 & 0.08468 & 0.00125 & $-0.22$ &  0.9 & 1,2,14,11 \\
GJ 892 & 21.141 & 0.290 & 4773 & 20 & 0.7784 & 0.0053 & 0.771 & 0.124 & 0.28046 & 0.00389 & $-0.23$ &  0.6 & 1,2,14,11 
\enddata
\tablecomments{Photometry references: (1) \citet{Skrutskie:2006lr}; (2) \citet{Morel:1978}; (3) \citet{Cousins:1980}; (4) \citet{Johnson:1965}; (5) \citet{Glass:1975}; (6) \citet{Koen:2010}; (7) \citet{Mould:1976}; (8) \citet{Johnson:1953}; (9) \citet{Persson:1977}; (10) \citet{Erro:1971}; (11) \citet{Johnson:1968}; (12) \citet{Bessel:1990}; (13) \citet{Johnson:1964}; (14) \citet{Johnson:1966}; (15) \citet{Cowley:1967}; (16) \citet{Veeder:1974}; (17) \citet{Weis:1993}; (18) \citet{Cutri:2003}; (19) \citet{Christou:2006}; (20) \citet{Eggen:1979}.}
\tablenotetext{a}{Errors on [Fe/H] are $\simeq0.07$~dex for all stars (since measurement errors are negligible compared to calibration errors).}
\tablenotetext{b}{$\chi^2_{\mathrm{red}}$ from our fit of the spectra to published photometry.}
\tablenotetext{c}{Some of the photometry for these stars comes from the General Catalogue of Photometric Data \citep{Mermilliod:1997qe}.}
\tablenotetext{d}{GJ 725B has a $\chi^2_\mathrm{{red}}$ of 7.4. The next highest $\chi^2_\mathrm{{red}}$ is 2.9 (GJ 725A), suggesting that either the photometry or spectroscopy for GJ 725B is inaccurate, or that the star is variable. For this reason we removed GJ 725B from our analysis.}
\label{tab:sample}
\end{deluxetable*}

We see no obvious cause for the disagreement between the photometric and spectroscopic data for GJ 725B. It is not active (no detectible H-$\alpha$ emission), and the fit for GJ 725A is significantly better ($\chi_{\mathrm{red}}^2$=2.9). Because our choice of which photometry to use for this star significantly affects our derived stellar parameters, we conservatively elect to remove this star from our calibration sample. 

\begin{figure}[htbp]
\begin{center}
   \includegraphics[width=0.45\textwidth]{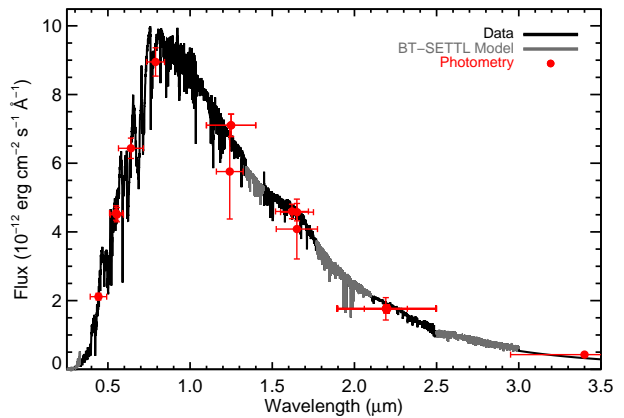} 
\caption{Spectrum of GJ 887 from SNIFS, SpeX, and PHOENIX models with photometry shown in red. Errors in the photometry are shown along the Y-axis, errors on the X-axis denote the approximate spectral region covered by the given filter. Regions of significant telluric contamination or gaps in the data are filled in using PHOENIX models, which are shown in gray. }
\label{fig:samplespectra}
\end{center}
\end{figure}

We show the observed spectrum and literature photometry of GJ 887 ($\chi^2_{\mathrm{red}}$ = 0.9) in Figure~\ref{fig:samplespectra} as a demonstration of our procedure. We report revised \fbol, \teff, and $L_*$ for the B12 sample in Table~\ref{tab:sample}. We include the derived stellar parameters for GJ 725B in Table~\ref{tab:sample}, although we caution readers that the errors in this star's parameters are probably underestimated.

\begin{figure}[htbp]
\begin{center}
   \includegraphics[width=0.45\textwidth]{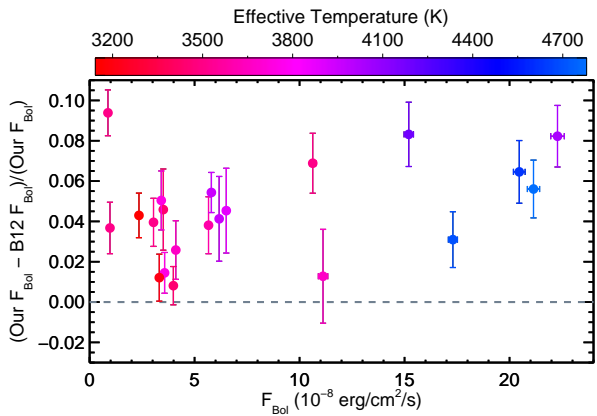} 
\caption{Fractional difference in our estimates of \fbol{} compared to those from B12. The dashed line indicates no difference. GJ 702B is not included because blended photometry prevents B12 from calculating an accurate \fbol{} for this star.}
\label{fig:fbolcomp}
\end{center}
\end{figure}

Our \fbol{} values are, on average, $\simeq$4\% higher than those reported by B12 (see Figure~\ref{fig:fbolcomp}), resulting in $\simeq$1\% higher \teff, and 4\% higher $\simeq L_*$. These differences are small, but typical errors in \fbol{} values are only $1--2\%$ so the offset is significant for many stars, and highly significant when considering the whole sample (see Figure~\ref{fig:fbolcomp}). 

This discrepancy is in part due to slightly different choices for input photometry between our study and that of B12. If we redo our analysis, but restrict ourselves to photometry reported in B12, the disagreement shrinks to $\simeq3\%$, mostly by moving in some of the largest outlier stars.  

\begin{figure}[htbp]
\begin{center}
   \includegraphics[width=0.45\textwidth]{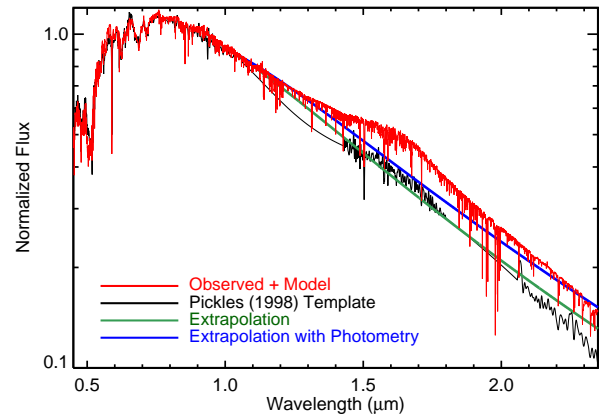} 
\caption{Observed spectrum (red) with the best-match \citet{1998PASP..110..863P} template star (black). We also show the extrapolation of the optical data into the NIR by fitting the 0.9-1.1\um\ region with a Planck function (green) and the fit if we include the NIR photometry (blue). }
\label{fig:pickles}
\end{center}
\end{figure}

We find that majority of the discrepancy can be explained by systematic issues with the \citet{1998PASP..110..863P} templates and extrapolation of these templates into the NIR. Of the 20 \citet{1998PASP..110..863P} templates K4 and later, only three of them have data past 1.1\um. These three the spectra are built by combining data from multiple sources, but still have gaps in each spectrum where the data do not overlap. We find that the combined templates do not properly conserve the relative flux calibration. An example can be seen in Figure~\ref{fig:pickles}, where we show a \citet{1998PASP..110..863P} template and our spectrum of GJ 380. Although the spectra are well matched in the optical, the NIR data are systematically offset. The result is differences in \fbol{} of 4\%--12\% between templates and our combined spectra. The situation improves if we `fix' the template by applying a constant offset to the template NIR data to better match our own spectrum, but offsets of 2\%--3\% in the derived \fbol{} values persist, similar to the difference between our own \fbol{} values and those from B12. 

The remaining 17 templates lack NIR data, and instead B12 must extrapolate beyond 1.1\um. We investigate how this affects the result by fitting our spectra in the 0.9-1.1\um\ region using a Planck function. The fit for GJ 380 is shown in Figure~\ref{fig:pickles}. Generally this method significantly underestimates the overall \fbol{} level (by 5\%--10\%). The discrepancy decreases if we include available NIR ($JHK$) photometry in the fit (resulting fit shown in Figure~\ref{fig:pickles}). However, even in this case the fit often fails to match the $H$-band ``bump'', resulting in discrepancies of 3\%--5\% in \fbol, which is similar to the offset between our results and that of B12. Late K and M dwarfs have spectra which are rich and irregular, and therefore cannot be fit with a simple function. 

\subsection{Masses}
B12 derive $M_*$ using the relation between $M_K$ and $M_*$ from \citet{Henry:1993fk}. They choose not to use the slightly more precise relation from \citet{2000A&A...364..217D} because the \citet{2000A&A...364..217D} sample includes only late K and M dwarfs, while B12 include many early K dwarfs. Because of our \teff{} cut, our calibration sample has only three stars that are outside the limits of the \citet{2000A&A...364..217D} relation. Therefore we derive $M_*$ using the relation from \citet{Henry:1993fk} for those three stars, and the \citet{2000A&A...364..217D} relation for the rest of the sample. Considering the conservative 10\% error assumed by \citet{2000A&A...364..217D} all $M_*$ values reported by B12 are within $1\sigma$ of our own. Resulting values of $M_*$ and errors are reported in Table~\ref{tab:sample}.

\subsection{Relations between \teff{} and fundamental parameters}\label{sec:relations}
\begin{figure*}[htbp]
\begin{center}
   \includegraphics[width=\textwidth]{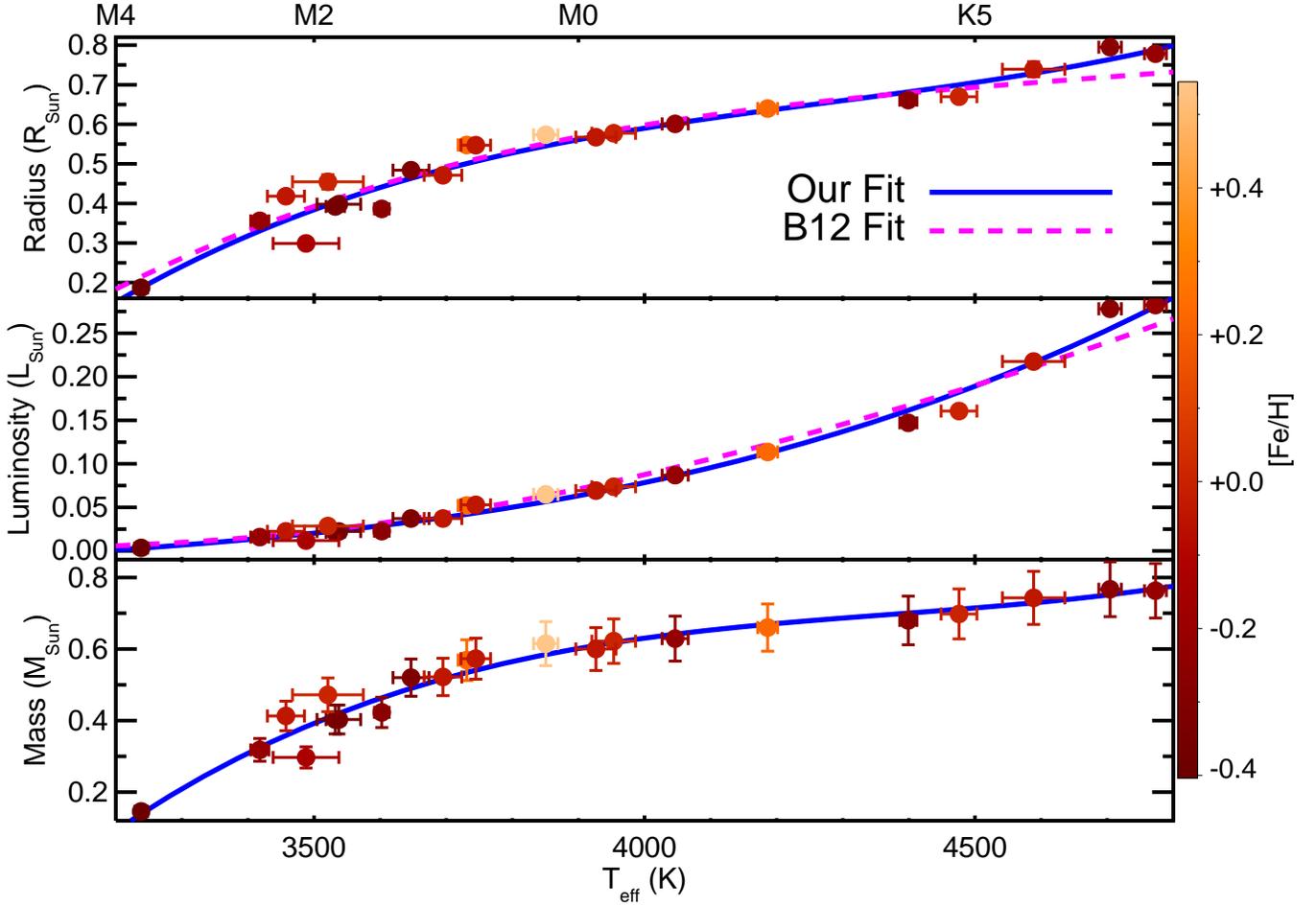} 
\caption{$R_*$, $L_*$, and $M_*$ of stars in our calibration sample as a function of their \teff. Stars are shaded (colored) based on their metallicities. Approximate spectral types are given for reference. Error bars are shown, although errors on $R_*$ and $L_*$ are sometimes smaller than the plotting points. Blue lines indicate our best-fit relations, described by Equations~\ref{eqn:RT}--\ref{eqn:RM}. The magenta dashed line shows the fits derived in B12 (B12 do not provide a \teff--$M_*$ relation).}
\label{fig:Tfits}
\end{center}
\end{figure*}

Our revision of \fbol, $T_{\mathrm{eff}}$, and $L_*$ values for the B12 stars changes the empirical relations B12 derive between \teff, $M_*$, $R_*$, and $L_*$. We use the Levenberg-Marquardt least-squares minimizer MPFIT \citep{2009ASPC..411..251M} to derive the following formulae:
\begin{eqnarray}
\label{eqn:RT} \nonumber R_* &=& -16.883175 + 1.1835396\times10^{-2}T_{\mathrm{eff}} \\  && - 2.70872\times10^{-6}T_{\mathrm{eff}}^2 + 2.105028\times10^{-10}T_{\mathrm{eff}}^3, \\
\label{eqn:RL} \nonumber L_* &=&  -0.78107639 + 7.3973011\times10^{-4}T_{\mathrm{eff}} \\  && - 2.4911\times10^{-7}T_{\mathrm{eff}}^2 + 2.947987\times10^{-11}T_{\mathrm{eff}}^3, \\
\label{eqn:RM} \nonumber M_* &=& -22.296508 + 1.5446387\times10^{-2}T_{\mathrm{eff}} \\  && - 3.488452\times10^{-6}T_{\mathrm{eff}}^2 + 2.64961\times10^{-10}T_{\mathrm{eff}}^3,
\end{eqnarray}
where \teff{} is given in Kelvin and $M_*$, $L_*$, and $R_*$ are given in solar units. Data and fits for these equations are shown in Figure~\ref{fig:Tfits} along with the corresponding fits from B12 (a \teff-$M_*$ relation is not shown because B12 do not derive a corresponding fit for comparison). The $\chi^2_\mathrm{{red}}$ for all three fits are $<1$, suggesting that some of the errors may be slightly overestimated. 

Errors for each fit are estimated using a Monte Carlo (MC) approach. Because of correlations between values (e.g., a change in \fbol{} alters \teff{} and $L_*$ in a coherent way) we perform our MC error estimate over all steps in our analysis. Specifically, we perturb the spectra, literature photometry, parallaxes, and angular diameters randomly according to their errors. We assume all sources of error are Gaussian, except for the additional error term in the SpeX data (see Section~\ref{sec:obs}), which affects the overall slope of the spectrum. Values of $M_*$ are assumed to have a correlated (but still Gaussian) error term, i.e., that derived $M_*$ for all stars may increase or decrease together for each MC run (due to possible systematics). This is treated separately from the non-correlated errors in the measurement of $M_K$ arising from errors in the photometry and parallax. 

\begin{figure}[htbp]
\begin{center}
   \includegraphics[width=0.45\textwidth]{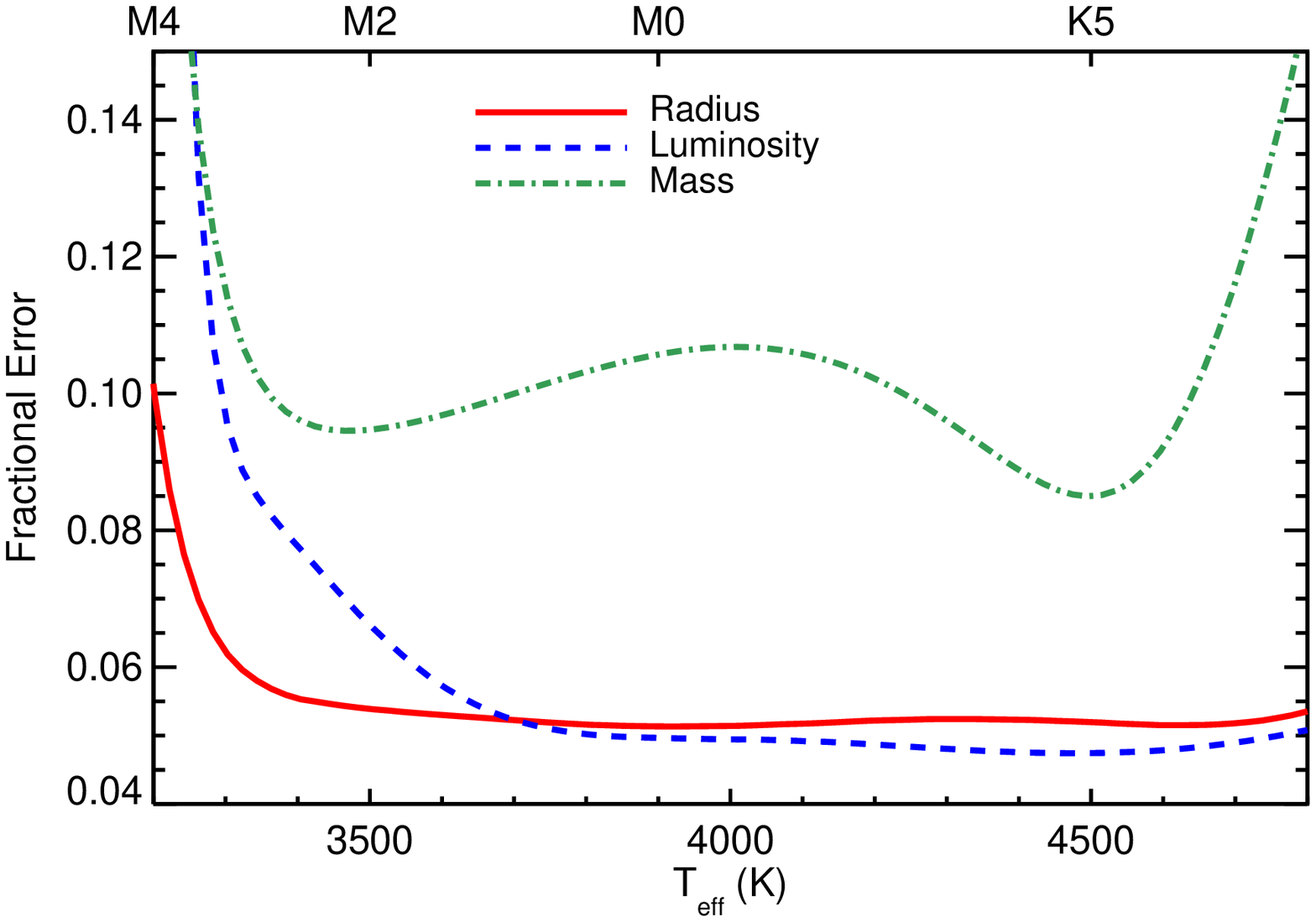} 
\caption{Fractional error as a function of \teff{} for Equations~\ref{eqn:RT}--ref{eqn:RM}. Approximate spectral types are shown for reference. Errors are calculated by a Monte Carlo process explained in Section~\ref{sec:relations} accounting for both the scatter in the data points around the fit and the uncertainty in the fit itself. Note that this error estimate does not include additional errors from measuring\ \teff. Rather, this represents the floor of what errors are possible using this technique.}
\label{fig:rel_errs}
\end{center}
\end{figure}

We perform the full analysis laid out in Section~\ref{sec:fixtab} using the perturbed values. In addition to perturbing the values we also randomly remove one calibrator star from the analysis (jackknifing), to account for sensitivity to our exact set of calibration stars. We repeat this process (perturbing values and randomly removing one star) 1000 times, each time we record the resulting polynomial fits between \teff, $R_*$, $M_*$, and $L_*$ and $L_*M_*^{-2/3}$. MC error estimation is also done for a relation between \teff{} and $L_*M_*^{-2/3}$ because this quantity is used in our HZ calculations (see Section~\ref{sec:habitable}), and errors in $L_*$ and $M_*$ are correlated. The adopted errors for each relation are assumed to be the standard deviation of the fitted $R_*$, $M_*$, and $L_*$ values at a given \teff. Errors as a function of \teff{} for each relation are shown in Figure~\ref{fig:rel_errs}. As expected, the errors grow significantly at \teff{} $<3300$~K where our calibration sample contains only a single star. 

\begin{figure}[htbp]
\begin{center}
   \includegraphics[width=0.45\textwidth]{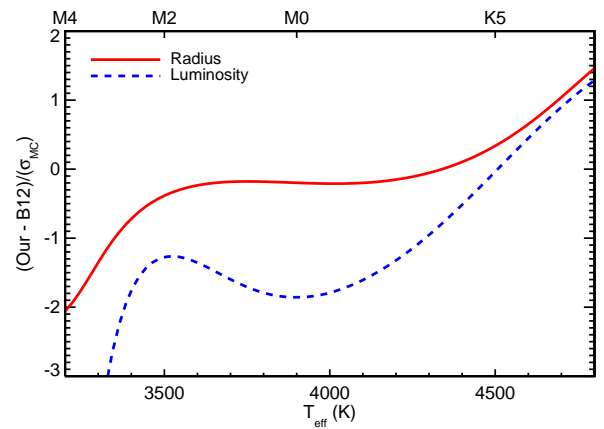} 
\caption{Difference between \teff-$R_*$ and \teff-$L_*$ relations from Equations \ref{eqn:RT}-\ref{eqn:RL} and those from B12 divided by the errors calculated from our MC/jackknife estimate (see Figure~\ref{fig:rel_errs}). Difference considers errors from our analysis only (not those reported by B12).}
\label{fig:rel_errs2}
\end{center}
\end{figure}

In Figure~\ref{fig:rel_errs2} we show the difference between our fits of $R_*$ and $L_*$ and those from B12 relative to our estimated errors. Note that for most temperatures, the differences are not significant. However, the offset is systematic and a function of temperature, and thus could be important. The higher discrepancy at cool temperatures (\teff$<3400$) is not due to our updated parameters, but instead due to the exclusion of GJ 725B from our data set. GJ 725B has an unusually large radius and luminosity for its temperature (assuming the assigned temperature is correct), which drives the B12 fits to larger radii and higher luminosities at low temperatures.

\subsection{Metallicity}\label{sec:metal}
B12 finds that radius is insensitive to metallicity, in contrast to model predictions \citep{Baraffe:1998kx, Dotter:2008fk}. However, such a conclusion requires accurate metallicities for late-type stars, which are notoriously difficult to determine. B12 draw metallicities for their K and M dwarfs from \citet{Edvardsson:1993qy}, \citet{Anderson:2011lr}, \citet{Rojas-Ayala:2012uq}, and \citet{2012A&A...538A..25N}. These sources use different techniques to determine metallicity, and likely suffer from systematic offsets from each other like those noted in \citet{Mann:2013gf}. Further, abundance analysis techniques from \citet{Edvardsson:1993qy} and \citet{Anderson:2011lr} are poorly calibrated for cool stars.

We estimate [Fe/H] for the 23 stars in our sample following the methods of \citet{Mann:2013gf}, which are tailored to data from SNIFS and SpeX. \citet{Mann:2013gf} provide empirical metallicity calibrations based on visible, $J$-, $H$-, and $K$-band spectra. We calculate the metallicity of each star using the weighted mean from each of the four calibrations, accounting for both measurement errors and systematic errors in the calibration as reported by \citet{Mann:2013gf}. 
 
\begin{figure}[htbp]
\begin{center}
   \includegraphics[width=0.45\textwidth]{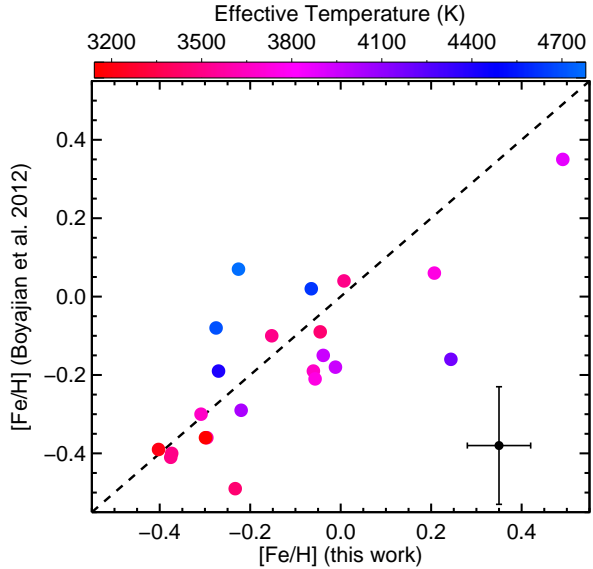} 
\caption{Comparison of [Fe/H] values derived from our analysis with those reported in B12 (see references within for original sources). A typical error bar is shown in the bottom right. Points are color-coded by their \teff. Note that errors in B12 metallicities vary based on the original literature source.}
\label{fig:metal}
\end{center}
\end{figure}

We report derived metallicities in Table~\ref{tab:sample}. In Figure~\ref{fig:metal} we show a comparison of our derived metallicities with those reported in B12. On average, our [Fe/H] values are 0.05~dex lower than those reported in B12. The difference in metallicity for 16 of the 23 stars (69\%) is $<1~\sigma$. Given the difficulty in measuring M dwarf metallicities, we consider the differences to be minor.
 
\begin{figure}[htbp]
\begin{center}
   \includegraphics[width=0.45\textwidth]{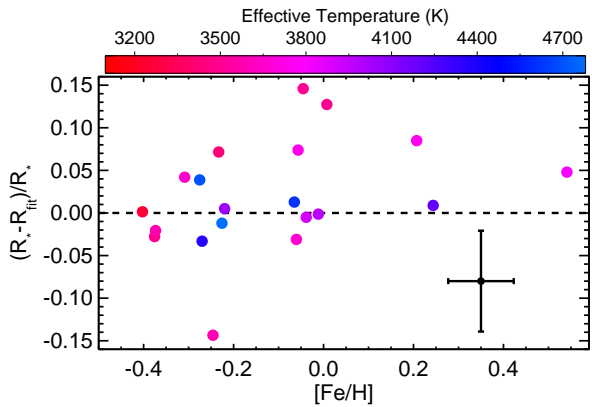} 
\caption{Fractional deviation of $R_*$ from the relation given by Equation~\ref{eqn:RT} as a function of [Fe/H]. Points are color-coded by their \teff. A typical error bar is shown in the bottom right. The error estimate includes errors arising from extrapolating \teff{} to $R_*$ (i.e., errors in \teff{} and errors in Equation~\ref{eqn:RT}). Although a weak correlation may be present, it is not statistically significant.}
\label{fig:metal_rad}
\end{center}
\end{figure}

Adding metallicity as a parameter in Equations~\ref{eqn:RT}--\ref{eqn:RM} does not improve the $\chi^2_{\mathrm{red}}$ in any significant way (for an example see Figure~\ref{fig:metal_rad}). The difference between a measured parameter (e.g., $R_*$) and the fitted value of that parameter shows no significant correlation with [Fe/H] for $M_*$, $R_*$, or $L_*$ based on the Spearman rank test (probability of a correlation $<70\%$ in all cases). M dwarf models predict a difference in the \teff--$R_*$ relation that should be detectable given the precision of our stellar parameters (see B12 for a further discussion of this discrepancy).

\section{Techniques to determine \teff}\label{sec:teff}
Equations~\ref{eqn:RT}--\ref{eqn:RM} provide a means to estimate other stellar properties from \teff. For nearby stars with parallaxes, we can determine \fbol{} from Equation~\ref{eqn:fbol} and $L_*$ from Equation~\ref{eqn:lum} and then constrain \teff{} by inverting Equation~\ref{eqn:RL}. However, we need a reliable means to measure \teff{} for stars without parallaxes (e.g., most \kepler stars). For this reason we use the calibration sample to develop and test techniques to measure \teff{} using primarily visible (but also NIR) spectra that can then be applied to more distant stars in a homogenous manner.

\subsection{Model Spectrum-fitting}\label{sec:modelfit}
We compare the SNIFS spectra of our calibration stars to a grid of models of K and M dwarf spectra from the BT-SETTL version of the PHOENIX code \citep{Allard:2011lr,Allard:2012fk}. We test the set of models that use the \citet{Asplund:2009zr} solar abundances (AGSS), and those using the updated \citet{Caffau:2011} revised solar abundances \citep[CIFIST,][]{Allard:2013, Rajpurohit:2013}. The AGSS models have been used extensively in previous studies of M dwarfs \citep[e.g.,][]{Mann:2012, Muirhead:2012pd, Lepine:2013lr}, but the CIFIST models use a revised TiO line list \citep{Plez:1998}. We download synthetic spectra for both model grids spanning \teff{} of 2700-5200~K in 100~K increments, log~$g$ values of 4.0, 4.5, and 5.0, and [M/H] values of $-1.0$, $-0.5$, 0.0, 0.3, and 0.5. [$\alpha$/Fe] is taken to be solar for [M/H]$\ge-0.5$ and +0.2 for [M/H]$\le-1.0$.

For a given stellar spectrum, we restrict our comparison to the subset of models with the two metallicity values immediately above and below the metallicity calculated for the star (see Section~\ref{sec:metal}). For example, for a star with [Fe/H] = 0.1 we restrict our comparison to models with [M/H] = 0.0 or [M/H] = +0.3. We assume the difference between [Fe/H] and [M/H] is negligible, which is reasonable for stars near solar metallicity. We place no restriction on models based on assumed log~$g$ or \teff{} other than those imposed by the range of grid points available.

Following \citet{Lepine:2013lr} we convolve the models with a Gaussian with full width at half maximum  of 7\AA\ ($R\simeq1000$) using the IDL code {\it gaussfold}. Although SNIFS spectra are already shifted to vacuum and rest frames, we correct for any additional wavelength differences by cross-correlating the observed and model spectra using the IDL routine {\it xcorl}. The size of this correction is typically less than one resolution element. 

M dwarf atmospheric models contain unknown systematic errors, in large part due to poorly modeled molecular absorption \citep[e.g., TiO bands,][]{Reyle:2011kx}. As a result, the distribution of (data-model)/error values will not be normally distributed and $\chi^2$ does not apply. Instead, we use the technique from \citet{Cushing:2008fj} and find the best-fit model by minimizing a goodness-of-fit statistic $G_K$, which is analogous to $\chi^2$. $G_K$ is defined as:
\begin{equation}
G_K = \sum_{i=1}^n \left( \frac{w_i(F_i-C_KF_{K,i})}{\sigma_i} \right)^2,
\end{equation}
where $w_i$ is the weight of the $i$th wavelength bin, $F_i$ and $F_{k,i}$ are the data and model flux densities, respectively, and $\sigma_i$ is the error on $F_i$. $C_K$ is a normalization constant that, for absolute flux calibrated spectra, is equal to $R_*^2/d^2$. Although we know $R_*^2/d^2$ for our calibration stars, neither the radius nor the distance are known generally. Instead we set $C_K$ such that the mean of $F$ and $F_K$ are the same.

Weights blueward of 5500\,\AA\ and redward of 9300\,\AA\ are set to 0 because of low S/N for fainter M dwarfs in the blue region and due to contamination from variable H$_2$O lines on the red end. H$_2$O variability is not a problem for the calibration stars because the exposure times are short (a few seconds), but this can be a major source of error for faint stars where the exposure times are often $\ge30$ minutes. 

\begin{figure}[htbp]
\begin{center}
   \includegraphics[width=0.45\textwidth]{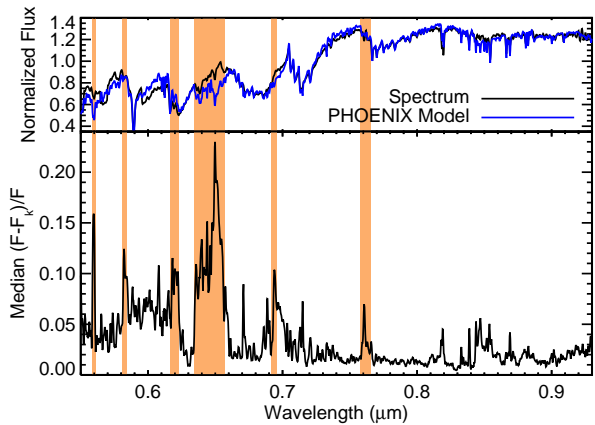} 
\caption{Top: sample spectrum (black) and best-fit PHOENIX (BT-SETTL) model (blue). Bottom: median fractional deviation of the 22 spectra from the best-fit PHOENIX model as a function of wavelength. Regions that are larger than 10\,\AA\ with a deviation larger than 10\% are masked out as part of our fitting process (shaded regions). The area around 0.76\um{} is excluded because of concerns about our ability to accurately remove the telluric O$_2$ line. Non-shaded regions are given full weight ($w_i=1$) in our model fits. }
\label{fig:residual}
\end{center}
\end{figure}

The rest of the weights are assigned either 0 or 1 based on how accurately models reproduce the real spectra. We first find the best-fit model for each of the calibration stars (described below) assuming $w_i=1$ for all $i$. We compute the residuals from the best-fit model for each star, and then compute the median fractional deviation between the data and model for all stars at a given wavelength. We show the final median spectral residuals in Figure~\ref{fig:residual}. From this stacked residual we identify regions with a median deviation of $>10\%$ ($>$ 1 order of magnitude greater than our measurement errors) that are larger than 10\,\AA\ wide. For these regions we set $w_i = 0$. All other weights are set to $w_i=1$.

We then repeat the process of fitting the model spectra to the data, computing the median residuals and adjusting the weights. After two iterations the weights do not change. We adopt this final list of weights for our model-fitting procedure. We also mask out a small region around 0.76\um{} due to concerns about accurate removal of the telluric O$_2$ line. We show the regions with $w_i=0$ (shaded) in Figure~\ref{fig:residual}. One part of the spectrum that stands out is the 0.64-0.66\um{} section, which contains the poorly modeled TiO absorption band \citep{Reyle:2011kx}. 

A more nuanced weighting scheme would be to weight each interval according to how consistent it is with the models, or test different weighting schemes to see which gives the best agreement with the bolometric temperatures. However regions with modest agreement between the real and synthetic spectra may contain more temperature information than regions with slightly better matches, and we have no a priori information about what spectral regions are the most temperature sensitive. Further, binary (0 or 1) weights are sufficient to get excellent agreement between the bolometric and model temperature scale.

After calculating $G_K$ for the relevant models, we select the seven models with the lowest $G_K$ values and construct 10,000 linear combinations from them \citep{Lepine:2013lr}. The \teff{} value assigned to each combination is the equivalent linear combination of the \teff{} values of the components. As before, we calculate $G_K$ for these 10,000 models. We adopt the temperature of the best-fit model from these linear combinations as the PHOENIX temperature (\teffbt).

\begin{figure}[htbp]
\begin{center}
   \includegraphics[width=0.45\textwidth]{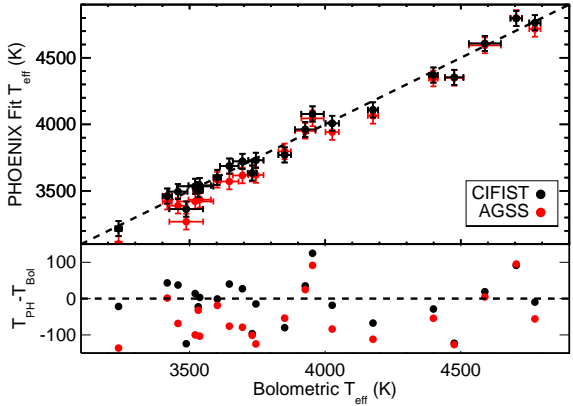} 
\caption{Comparison of \teff{} determined from fitting PHOENIX models (AGSS and CIFIST) to visible-wavelength spectra to those calculated from bolometric values calculated with Equation~\ref{eqn:teff_fbol}. Red points correspond to fits from the AGSS models, while black points correspond to fits from the CIFIST models. The bottom plot shows the residuals assuming a 1:1 relation. The dashed line indicates equality (for both plots). The rms deviation of \teffbt{} from \teffbol{} is when using the CIFIST models is 57~K, which we adopt as the error on our model spectrum-fitting procedure.}
\label{fig:tcomp}
\end{center}
\end{figure}

We compare \teffbt{} from each model grid (AGSS and CIFIST) to the \teff{} derived from bolometric flux and angular diameter (\teffbol) in Figure~\ref{fig:tcomp}. Using either grid \teffbt{} accurately reproduces \teffbol{} for stars warmer than 4000~K. For cooler stars the AGSS models predict significantly cooler temperatures than those derived bolometrically. We find a best-fit linear relation of:
\begin{equation}\label{eqn:corr}
T_{\mathrm{eff,BOL}} = 279.(\pm43) + 0.942(\pm0.013)T_{\mathrm{eff,AGSS}}.
\end{equation}
Errors (quoted in parentheses) are determined via MC. Specifically, we randomly perturb \teffbol{} values and the visible-wavelength spectra by reported errors. We then re-fit the spectra using by minimizing $G_K$ as explained above. We repeat this 1000 times, and report the 1-$\sigma$ errors. The difference between \teffbt{} and \teffbol{} is small for warm stars, but results in a systematic offset of more than 75~K at 3500~K. We find that the slope of \teffbt{} vs. \teffbol{} using the AGSS models is inconsistent with 1 at 5$\sigma$, and conclude that this slope cannot be due to random noise. If, instead, we adopt the B12 temperatures for our calibration stars (but still use the AGSS models) we derive the relation:
\begin{equation}\label{eqn:corr}
T_{\mathrm{eff,B12}} = 366.(\pm41) + 0.907(\pm0.011)T_{\mathrm{eff,AGSS}}.
\end{equation}
Thus the difference between temperatures derived using the AGSS models has a more significant slope (8$\sigma$) when we use the B12 temperatures.

We note a significant slope between CIFIST and B12 temperatures. We derive a best fit relation of:
\begin{equation}\label{eqn:corr}
T_{\mathrm{eff,B12}} = 192.(\pm41) + 0.960(\pm0.012)T_{\mathrm{eff,CIFIST}}.
\end{equation}
Although this slope is still 4$\sigma$ inconsistent with 1, CIFIST models are still better at reproducing the B12 temperature scale than AGSS models.

Effective temperatures derived using the CIFIST grid are in excellent agreement with our derived \teffbol values, with a median difference of 1 ($\pm13$)~K, and no statistically significant slope (0.99$\pm0.01$), as can be seen in Figure~\ref{fig:tcomp}. The rms deviation of \teffbt{} from \teffbol{} when using the CIFIST models is just 57~K (better than when using AGSS even after correcting for the offset), which we adopt as the error from our model fitting procedure. Because of the superior performance of the CIFIST models in reproducing our \teffbol{} scale (and even the B12 \teffbol{} scale) we use only CIFIST models for the rest of the paper.

\subsection{Temperature Sensitive Indices}\label{sec:indices}
In addition to our model spectrum-fitting procedure we develop four novel temperature indices, one for each wavelength regime (visible, $J$-, $H$-, and $K$-band). We define an index as: 
\begin{equation}\label{eqn:index}
\mathrm{Index} = \frac{<\mathrm{Band1}>/<\mathrm{Band2}>}{<\mathrm{Band2}>/<\mathrm{Band3}>},
\end{equation}
where $<>$ denotes the median flux level within a given wavelength region. An index is a measure of the curvature of the spectrum in a particular region. This definition is identical to that given in \citet{2010ApJ...722..971C} although they define bands to measure the level of H$_2$O absorption in NIR spectra of M dwarfs, while we are interested in an empirical method to measure \teff{} with no preference for any particular absorption band. 

We search for the best performing (lowest $\chi^2$) band definitions for each wavelength regime: visible (0.3 - 1.0\um), $J$- (1.0 - 1.45\um), $H$- (1.45 - 1.85\um), and $K$-band (1.85 - 2.4\um). We vary the central wavelengths of each band ($\lambda_{c,1}$, $\lambda_{c,2}$, $\lambda_{c,3}$) in increments of 5\,\AA, and the width of the bands ($\delta \lambda$) from 25\,\AA{} to 100\,\AA{} in increments of 5\,\AA. We exclude bands which would overlap with features that are identified by \citet{Mann:2013gf} as significantly metal sensitive.

For every iteration we calculate the index (Equation~\ref{eqn:index}) for each calibration star. We fit a second-order polynomial between the index measurement and the \teffbol{} values and calculate $\chi^2$. The best $\lambda_i$ and $\delta \lambda$ values are taken to be the ones that give the smallest $\chi^2$. We use the final rms of the fitted temperature vs. the \teffbol{} as an estimate of the error for each relation.

\begin{figure*}[htbp]
\begin{center}
   \includegraphics[width=\textwidth]{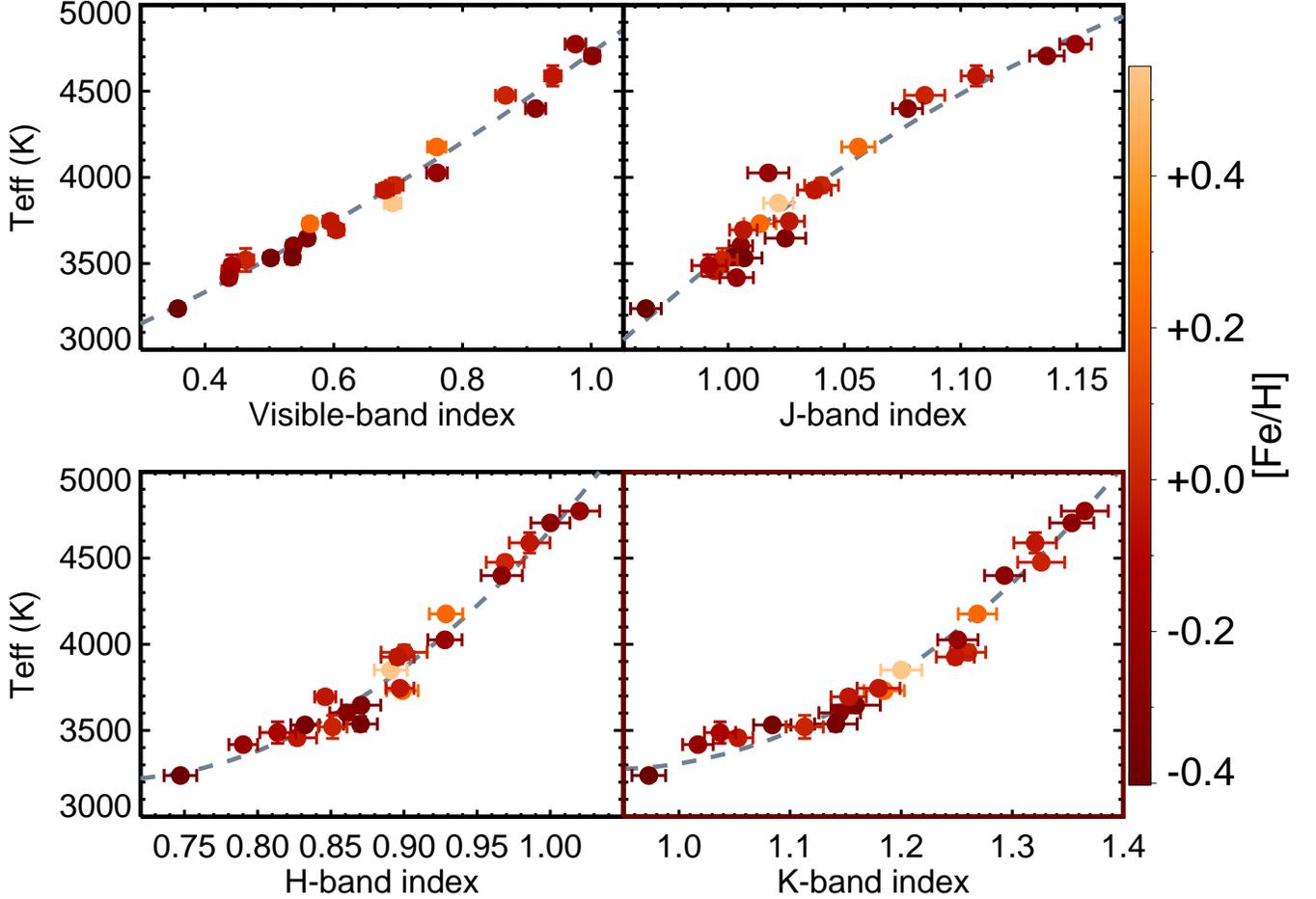} 
\caption{Relation between our empirical indices and \teff{} for the B12 sample. See Table~\ref{tab:indices} for index definitions. Note that many errors are smaller than the data points. Each point is colored according to its metallicity (see legend) to demonstrate that there is no obvious trend with metallicity. }
\label{fig:indices}
\end{center}
\end{figure*}

We report the $\lambda_i$ and $\delta \lambda$ values, the polynomial coefficients for calculating \teff{} from these indices, and the adopted errors from each relation in Table~\ref{tab:indices}. We also show the indices and \teff{} values for each wavelength regime in Figure~\ref{fig:indices}. The relation for visible-wavelength spectra gives the smallest error, although it is still not as precise as the model spectrum-fitting technique. Most likely the model spectrum-fitting procedure performs better because it is using more spectral information. However, these relations have the advantage of being model independent, and can be calculated easily from low or moderate-resolution spectra. 

\begin{deluxetable*}{l | l l l l l l l l r r r}
\tablecaption{Index Band Definitions:\\ \teff{} = $a$ + $b$(Index) + $c$(Index$^2$)}
\tablewidth{0pt}
\tablehead{
\colhead{Name} & \colhead{Coverage} & \colhead{$\lambda_{c,1}$} & \colhead{$\lambda_{c,2}$} & \colhead{$\lambda_{c,3}$} & \colhead{$\delta \lambda$} & \colhead{$a$} & \colhead{$b$} & \colhead{$c$} & \colhead{$\sigma_{T_{\mathrm{eff}}}$} \\
\colhead{} & \colhead{\um} & \colhead{\um} & \colhead{\um} & \colhead{\um} & \colhead{\AA} & \colhead{} & \colhead{} & \colhead{} & \colhead{K} 
}
\startdata
$V$ & 0.52 - 0.95 & 0.5700 & 0.6500 & 0.6850 & 30 & 2.683$\times10^3$ & 1.354$\times10^{3}$ & 6.850$\times10^{2}$ & 62 \\
$J$ & 1.00 - 1.45 & 1.0130 & 1.1010 & 1.1630 &  85 & $-2.277\times10^{4}$ & 4.199$\times10^{4}$ & $-1.565\times10^{4}$ & 101\\
$H$ & 1.45 - 1.81 & 1.4640 & 1.6630 & 1.7960 &  30 & 1.087$\times10^{4}$ & $-2.196\times10^{4}$ & 1.568$\times10^{4}$ &  80\\
$K$ & 2.00 - 2.35 & 2.0260 & 2.2765 & 2.3385 &  35 & 1.051$\times10^{4}$ & $-1.543\times10^{4}$ & 8.228$\times10^{3}$ & 73 
\enddata
\tablecomments{Index$_i$ is defined by Equation~\ref{eqn:index} using the bands defined in this table. $\lambda_{c,i}$ denotes the central wavelength of a given band in \um, $\delta \lambda$ denotes the width of a given band in \AA.}
\label{tab:indices}
\end{deluxetable*}

Screening out metal--sensitive regions from \citet{Mann:2013gf} reduces the chance of systematics with metallicity in our \teff-index relations. However, \citet{Mann:2013gf} detects metal sensitive atomic and molecular lines, and do not investigate the role of metallicity on continuum shape. To test the effect of metallicity on our indices we perform a Spearman rank test on \teff{} - $T_{\mathrm{eff,\,fit}}$ vs. [Fe/H] for all relations. We find a correlation probability of 5\%, 9\%, 20\%, and 66\% for the visible, {$J$-,} {$H$-,} and {$K$-band} relations, respectively. We conclude that metallicity is a minor effect on these indices for the metallicity regime covered by our sample.

\section{Application to \kepler planet hosts}\label{sec:kepler}
We apply our methods to calculate the \teff, $M_*$, $R_*$, and $L_*$ of the sample of \kepler planet hosts described in Section~\ref{sec:sample}. We use our model spectrum-fitting procedure (using the CIFIST models) to determine \teff{} because this gives the most precise \teff{} values. One immediate concern is that the S/N of the visible spectra for our calibration sample is extremely high (S/N\,$>200$) compared to that of the \kepler target sample (S/N$\simeq$80). We test the effect of lower S/N by adding Gaussian noise to the spectra of our calibration sample. We then repeat our process described in Section~\ref{sec:modelfit}. We find that the rms is essentially unchanged ($\sigma_{T_\mathrm{eff}} = 58$ instead of $\sigma_{T_\mathrm{eff}} = 57$) and no significant systematic offset is present at the S/N of the \kepler target sample. Only at S/N$\le50$ does the error in our model spectrum-fitting procedure begin to rise steeply, most likely because above this S/N systematic errors dominate over measurement errors. 

We report \teff, $M_*$, $R_*$, and $L_*$ values for each of the late-type \kepler planet hosts in Table~\ref{tab:kepler}. Quoted errors consider both errors in Equations (5)-(7) (modeled by MC) and errors arising from our measurement of \teff. 

\begin{deluxetable*}{l l l l l l l l l l l}
\tablecaption{Late-type \kepler planet host parameters}
\tablewidth{0pt}
\tablehead{
\colhead{KOI \#} & \colhead{KID \#} & \colhead{\teff{}\tablenotemark{a}} & \colhead{$R_*$} & \colhead{$\sigma_{R}$} & \colhead{$M_*$} & \colhead{$\sigma_{M}$} & \colhead{$L_*$} & \colhead{$\sigma_{L}$}  \\
\colhead{} & \colhead{} & \colhead{K} & \colhead{$R_\odot$} & \colhead{$R_\odot$} & \colhead{$M_\odot$} & \colhead{$M_\odot$} & \colhead{$L_\odot$} & \colhead{$L_\odot$} 
}
\startdata
 227 &    6185476 & 4093 &  0.615 &  0.035 &  0.653 &  0.070 &  0.095 &  0.012\\
 247 &   11852982 & 3852 &  0.547 &  0.034 &  0.586 &  0.065 &  0.057 &  0.008\\
 248 &    5364071 & 3970 &  0.583 &  0.034 &  0.623 &  0.068 &  0.074 &  0.010\\
 249 &    9390653 & 3548 &  0.413 &  0.039 &  0.428 &  0.058 &  0.024 &  0.005\\
 250 &    9757613 & 4049 &  0.604 &  0.034 &  0.643 &  0.070 &  0.087 &  0.011\\
 251 &   10489206 & 3770 &  0.517 &  0.035 &  0.553 &  0.062 &  0.047 &  0.007\\
 254\tablenotemark{b} &    5794240 & 3820 &  0.536 &  0.034 &  0.574 &  0.064 &  0.053 &  0.008\\
 314\tablenotemark{c} &    7603200 & 3871 &  0.553 &  0.034 &  0.593 &  0.065 &  0.060 &  0.008\\
 940\tablenotemark{d} &    9479273 & 5085 &  0.938 &  0.076 &  0.885 &  0.284 &  0.416 &  0.038\\
 952\tablenotemark{e} &    9787239 & 3801 &  0.529 &  0.035 &  0.566 &  0.063 &  0.051 &  0.008\\
 961\tablenotemark{f} &    8561063 & 3241 &  0.189 &  0.055 &  0.143 &  0.073 &  0.003 &  0.003\\
1361\tablenotemark{g} &    6960913 & 4158 &  0.630 &  0.035 &  0.665 &  0.070 &  0.107 &  0.013\\
2704\tablenotemark{h} &    9730163 & 3157 & $<$0.187 & \nodata & $<$0.159 & \nodata & $<$0.0034 & \nodata 
\enddata
\tablecomments{(This table is available in its entirety in a machine-readable form in the online journal. } 
\tablenotetext{a}{Error on \teff{} 58~K for all stars (see Section~\ref{sec:modelfit}).}
\tablenotetext{b}{Consistent with \teff{} = 3820$\pm$90~K $R_*=0.55\pm0.11$ and $M_* = 0.59 \pm 0.06$ from \citet{Johnson:2012fk}.}
\tablenotetext{c}{Consistent with $R_* = 0.54 \pm 0.05$, $M_* = 0.57 \pm 0.05$ from \citet{Pineda:2013fk}.}
\tablenotetext{d}{KOI 940 and KOI 2704 have \teff{} values outside our grid of calibration stars.}
\tablenotetext{e}{Kepler-32. Consistent with \teff{} = $3793^{+80}_{-74}$, $R_* = 0.53 \pm 0.02$, $M_* = 0.54 \pm 0.02$ from \citet{Swift:2013vn}.}
\tablenotetext{f}{Kepler-42. Consistent with \teff{} = $3068\pm+174$, $R_* = 0.17 \pm 0.04$, $M_* = 0.13 \pm 0.05$ from \citet{Muirhead:2012zr}.}
\tablenotetext{g}{Kepler-61. Consistent with \teff{} $= 4017^{+68}_{-150}$, $R_*=0.62^{+0.02}_{-0.05}$, and $M_* = 0.635 \pm 0.037$ from \citet{Ballard:2013}.}
\tablenotetext{h}{KOI 2704 has a \teff{} cooler than our grid of calibration stars. We conservatively assign upper limits to this star based on the calibration star with the best matched temperature (GJ 699).}
\label{tab:kepler}
\end{deluxetable*}

We derive a \teff{} of 5071$\pm$57~K for KOI 940, which is significantly warmer than the next warmest star (KOI 2417, \teff{} = 4582$\pm57$), and makes it an early K dwarf. This temperature is outside the range spanned by our calibration sample, and thus the derived stellar parameters should be viewed skeptically. KOI 940 makes our $K_P-J$ color cut (Section~\ref{sec:sample}) most likely due to an erroneous $K_P$ magnitude, or because of significant reddening. This illustrates a drawback of using $K_P-J$ to define our sample instead of the full spectral energy distribution.

The coolest star in the sample is KOI 2704 with a \teff{} of 3157$\pm$57~K. This star is likely even cooler than KOI 961 (Kepler-42, \teff{} = 3241$\pm57$). Unfortunately, KOI 2704's \teff{} is slightly below the \teff{} range of our calibrators (the coolest is GJ 699 with a \teff{} of 3238$\pm$ 11), which makes it difficult to constrain its physical parameters using the methods outline in this paper. For example, our empirical relations yield a $L_*$ below 0 for this \teff, which is obviously unphysical. Instead, we conservatively assign 1$\sigma$ upper limits to the parameters of this star corresponding to the coolest calibrator star in the sample (GJ 699). Additional, cooler calibration stars are needed to better constrain the parameters of this star.

\begin{figure}[htbp]
\begin{center}
   \includegraphics[width=0.45\textwidth]{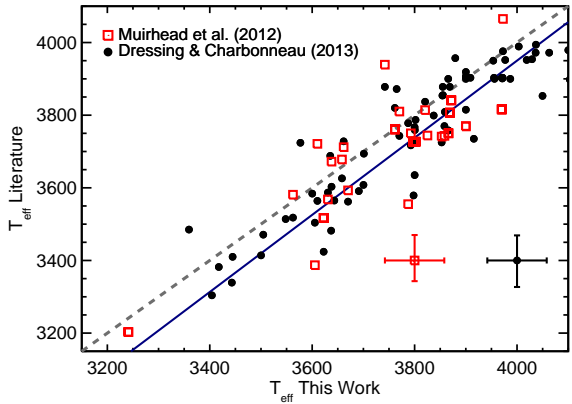} 
\caption{\teff{} values from \citet{Dressing:2013fk} (black circles) and \citet{Muirhead:2012pd} (red squares) vs. those found in this work. The gray dashed line indicates equality. The solid blue line denotes equality if we used the AGSS models instead of CIFIST (see Section~\ref{sec:modelfit}). Typical error bars are shown in the bottom right color-coded by their corresponding source.}
\label{fig:teffcomp}
\end{center}
\end{figure}

We show a comparison of our \teff{} values to those from \citet{Muirhead:2012pd} and \citet{Dressing:2013fk} in Figure~\ref{fig:teffcomp}. Because the H$_2$O index is only sensitive to \teff{} below $4000$~K \citep{Rojas-Ayala:2012uq,Dressing:2013fk,Mann:2013vn}, we exclude stars warmer than $4000$~K from \citet{Muirhead:2012pd} in our comparison sample. Our median \teff{} is $43^{+12}_{-8}$ higher than those from \citet{Dressing:2013fk} and $72^{+9}_{-12}~K$ higher than those from \citet{Muirhead:2012pd} (errors based on bootstrap resampling).

Most of this disagreement can be explained by the models utilized by the different groups. \citet{Muirhead:2012pd} \teff{} values are based on a calibration of the H$_2$O-K2 index to AGSS models \citep{Rojas-Ayala:2012uq}. \teff{} from \citet{Dressing:2013fk} are based on matching photometry from the {\it Kepler} Input Catalog \citep{Brown:2011fj} and the Two-Micron All Sky Survey \citep{Skrutskie:2006lr} to \citet{Dotter:2008fk} models, which are is based on an earlier version of the PHOENIX code that utilizes the \citet{1998SSRv...85..161G} solar abundances \citep{1999ApJ...512..377H,1999ApJ...525..871H}. Had we used AGSS models in our model-fitting procedure, the median differences would have been $0^{+7}_{-7}~K$ and $6^{+6}_{-8}$ (both consistent with no difference) for \citet{Dressing:2013fk} and \citet{Muirhead:2012pd}, respectively.

\section{Planets in the Habitable Zones of \emph{M Dwarfs}}\label{sec:habitable}

Revised estimates of \teff{}, $R_*$, and $L_*$ allow us to determine the irradiance, $S$, and radius, $R_P$, of the 188 planets around these stars and determine their position with respect to the HZs of their host stars \citep{Dressing:2013fk,Kopparapu2013b,Gaidos:2013rt}.  We estimated $S$ experienced by each of the 191 KOIs around these late-type dwarfs using the $L_*$ values above, the orbital periods as from {\it Kepler}, and the $M_*$ from our effective \teff-$M_*$ relation (see Section~\ref{sec:relations}). Assuming near-circular orbits, the orbit-averaged $S$ in terrestrial units is given by:
\begin{equation}
\frac{S}{S_{\oplus}} \simeq \frac{L_*}{L_{\odot}}\left(\frac{P}{\rm 365~\mathrm{days}}\right)^{-4/3}\left(\frac{M_*}{M_{\odot}}\right)^{-2/3}.
\end{equation}

\begin{figure}
   \includegraphics[width=0.45\textwidth]{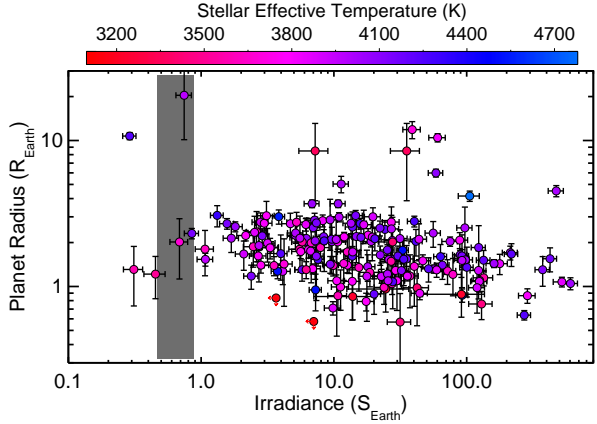} 
\caption{$R_P$ of \kepler planets around M dwarf stars vs. estimated $S$ (both in terrestrial units). Points are color-coded by their stellar \teff{} following the same color scheme as Figures~2, 5, and 6. The $R_P$ and $S$ values of KOIs 2704.01 and 2704.02 are assigned upper limits because the late M host stars is outside our range of calibrated stellar parameters. The bounds of the HZ for a \teff=3800~K star are shown as a shaded region \citep{Kopparapu2013}.}
\label{fig:rad_irr}
\end{figure}

We estimated the planet radius using $R_P/R_*$ from \citet{Dressing:2013fk}, where available, and from the KOI catalog in the MAST database\footnote{http://archive.stsci.edu/kepler/koi/search.php}, otherwise. Figure~\ref{fig:rad_irr} shows $R_P$ vs. $S$ with the points color-coded by the \teff{} of the host star.  Errors in radius account for the errors in $R_*$ and $R_P/R_*$; the latter are taken from MAST as \citet{Dressing:2013fk} do not report these values. Errors in $R_P/R_*$ dominate for the smaller planets.  Estimated $R_P$ and $S$ are given in Table~\ref{tab:planet_properties}.

\begin{deluxetable}{l l l l l l l l l l l l l }
\tablecaption{\kepler planet parameters}
\tablewidth{0pt}
\tablehead{
\colhead{KOI \#} & \colhead{KID \#} & \colhead{$R_P$} & \colhead{$\sigma_{R_P}$} & \colhead{$S$} & \colhead{$\sigma_S$}  \\
\colhead{} & \colhead{} & \colhead{$R_\earth$} & \colhead{$R_\earth$} & \colhead{$S_\earth$} & \colhead{$S_\earth$}}
\startdata
  227.01 & 6185476 & 2.853 & 0.114 & 7.151 & 0.897\\
  247.01 & 11852982 & 1.791 & 0.127 & 6.421 & 0.900\\
  248.01 & 5364071 & 2.036 & 0.083 & 19.0 & 2.5\\
  248.02 & 5364071 & 2.991 & 0.072 & 10.9 & 1.5\\
  248.03 & 5364071 & 2.036 & 0.073 & 74.9 & 10.0\\
  248.04 & 5364071 & 2.163 & 0.083 & 5.373 & 0.717\\
 2704.01\tablenotemark{a} & 9730163 & 0.835 & 0.667 & 3.7 & \nodata \\
 2704.02\tablenotemark{a} & 9730163 & 0.576 & 0.672 & 7.0 & \nodata 
\enddata
\tablecomments{(This table is available in its entirety in a machine-readable form in the online journal.)} 
\tablenotetext{a}{KOI 2704 has a \teff{} below our coolest calibration star. Therefore only upper limits are given.}
\label{tab:planet_properties}
\end{deluxetable}

We adopt a conservative definition of the HZ as the range of $S$ bracketed by the runaway greenhouse limit and the first CO$_2$ condensation limit for an Earth-like atmosphere on a one Earth-mass ($M_\earth$) planet.  We use the empirical relations described in \citet{Kopparapu2013}, corrected in \citet{Kopparapu2013Err} and available in the accompanying online calculator. These calculations do not account for water clouds other than a decrease in the effective planetary albedo.  Moreover, these criteria will not apply to planets with a very un-Earth-like atmosphere \citep{Pierrehumbert2011} or much more massive planets where the surface gravity (and hence pressure, for the same optical depth) is much higher.

\begin{figure}
   \includegraphics[width=0.45\textwidth]{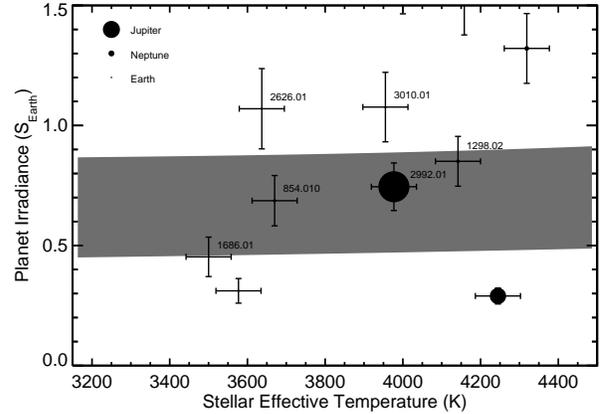} 
\caption{Estimated $S$ of \kepler planets around M dwarf stars (terrestrial units) vs. the \teff{} of the host star.  The upper and lower bounds of the HZ (shaded region) as defined by the runaway greenhouse and first CO$_2$ condensation limits are calculated based on \citet{Kopparapu2013}.  Data points are scaled in size by the estimated planet size. For the smallest planets the plotted points are smaller than the thickness of the error bars, while for the largest planets the error bars are smaller than the point. Objects within 2$\sigma$ of the HZ are labeled. Most planets are off scale above the top of the plot.}
\label{fig:hz}
\end{figure}

Figure \ref{fig:hz} plots $S$ (in terrestrial units) vs. \teff{} for planets near or in the HZ.  Because estimates of the $L_*$ and $M_*$ of stars are covariant, errors in the quantity $L_*M_*^{-2/3}$ were calculated by MC (Section \ref{sec:relations}).  We ignore errors in $P$, which are comparatively small.  Four candidate planets (KOIs 854.01, 1298.02, 1686.01, and 2992.01) lie within or within one standard deviation of the HZ. Based on the $R_P/R_*$ from MAST, we estimate the radius of KOI 2992.01 to be $\simeq$20R$_\earth$ and thus would not be a planet. However, the impact parameter given in MAST is 1.28, suggesting possible problems with the analysis of the \kepler light curve. Our upper limit on the $L_*$ of the late M dwarf system KOI 2704 means that either or both of two of its candidate planets (components 0.01 and 0.02) might be in the HZ, although we are unable to make reliable assignments at this time. Two other candidate planets (KOI 2626.01 and KOI 3010.01) lie marginally (1.2 and 1.5$\sigma$) interior to the HZ.  

We find that two of the four Earth-size (0.5-1.4$R_\earth$) planets identified by \citet{Kopparapu2013b} as orbiting in the HZ are actually outside the HZ: KOIs 1422.02, 2418.01, and 2626.01. Although KOI 2626.01 is only 1.2$\sigma$ outside the HZ, KOI 1422.02 is significantly (4.1$\sigma$) interior to the HZ. A third planet, KOI 2418.01, is significantly (2.7$\sigma$) outside the CO$_2$ condensation definition of the limit of the HZ, but falls within the `maximum CO$_2$ greenhouse definition of the outer limit used by \citet{Kopparapu2013b}. These differences are primarily due to our revised estimates of \teff{} and indicate the importance of accurate \teff{} estimates for stars without parallaxes. Removal of these two objects from the tally of Earth-size HZ planets (our additions of 854.01, 1298.02, and 2992.01 have radii of $\ge2R_\earth$) reduces the \citet{Kopparapu2013b} estimated value of $\eta_{\oplus}$. 

\begin{figure}
   \includegraphics[width=0.45\textwidth]{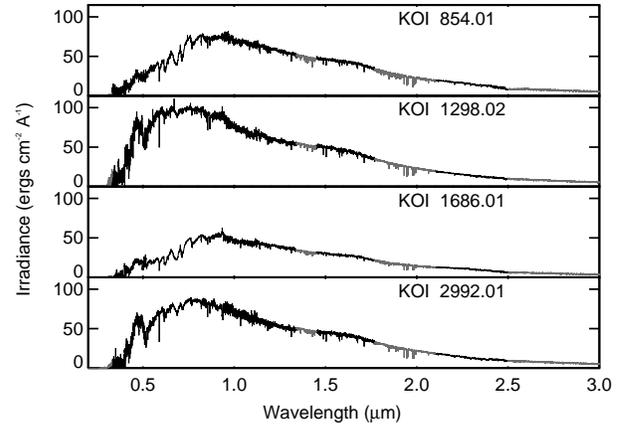} 
\caption{Spectra of the stellar irradiance experienced by four
  candidate planets in (or within 1$\sigma$ of the HZ, scaled using the estimated
  $L_*$ of each host star.  Gray points interpolate across
  observationally inaccessible wavelength regions using the best-fit
  PHOENIX model.}
\label{fig:spectra}
\end{figure}

Figure \ref{fig:spectra} plots the stellar $S$ spectra experienced by the four candidate HZ planets, scaled using the estimated $L_*$.  We interpolate across gaps in wavelength regions outside the observationally accessible windows using the best-fit PHOENIX models (gray regions).

\begin{figure}
   \includegraphics[width=0.45\textwidth]{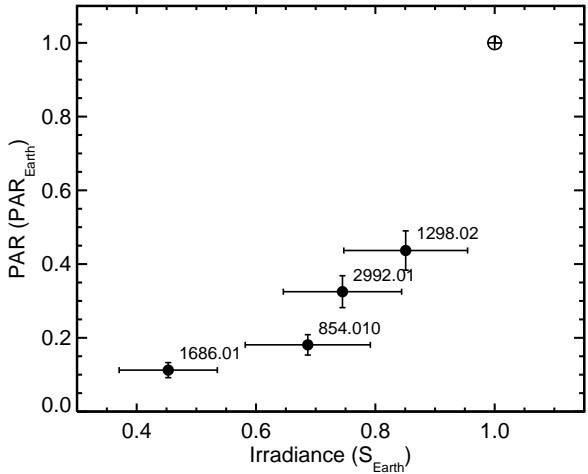} 
\caption{Photosynthetic active radiation (top of the atmosphere) in terrestrial units vs. total irradiation (top of the atmosphere) in terrestrial units for \kepler planet candidates in the HZs of late K and early M dwarfs. Earth is shown for reference.}
\label{fig:par}
\end{figure}

The spectra of M dwarfs are characteristically less intense at blue wavelengths relative to the Sun.  On a planet in the HZ of an M dwarf there would be comparatively less photosynthetic active radiation (PAR, 400-700~nm) available for hypothetical life capable of oxygenic photosynthesis. \citet{2002Icar..157..535W} show that a planet at the inner edge of the HZ of an M0 dwarf would receive about one third of PAR photons compared to an equivalent planet around a G2 dwarf.  Using our spectra we calculated the incident normal flux in PAR received at the orbit of each candidate HZ planet. We normalized by the total solar irradiation in PAR (530.1 W m$^{-2}$) based on the 2000 ASTM Standard Extraterrestrial Spectrum REference E-490-00\footnote{http://rredc.nrel.gov/solar/spectra/am0/ASTM2000.html}. We find that these candidate HZ planets receive anywhere from 10\% to 45\% of the PAR as the Earth (Figure~\ref{fig:par}); this varies somewhat from the of \citet{2002Icar..157..535W} in part because these stars cover a range of spectral types. However, lest this dim prospects for plants on M dwarf planets, the reduced PAR is still orders of magnitude higher than the minimum light requirement of terrestrial oxygen-evolving photosynthetic life \citep{Cockell2009}.  Indeed, operation of the photosystem II moiety of oxygenic photosynthesis can be inhibited by high light conditions \citep[photoinhibition,][]{Long:1994eu}, a phenomenon which would be relieved in the HZ of an M dwarf.  Moreover, two chlorophylls ($d$ and $f$) have been recently identified as capable of harvesting light as red as 750~nm for oxygenic photosynthesis \citep{Chen2011}.  Such adaptations may be important for life around a red dwarf.

\section{Discussion}\label{sec:discussion}
Making use of nearby, well-characterized K and M dwarfs, we develop and calibrate techniques to determine the physical characteristics of late-type dwarfs from moderate-resolution spectra (1000\,$\le R\le$\,2000). Our method is most effective for faint and/or distant stars lacking parallaxes, and is observationally inexpensive compared to techniques that require high-resolution spectra \citep[e.g.,][]{Pineda:2013fk}. 

As part of our analysis, we take spectra of and re-derive \teff, $M_*$, and $L_*$ for a sample of stars with interferometrically measured radii. Our spectra of these stars are available online. Our derived \fbol{} values are systematically higher than those previously estimated by B12 (Figure~\ref{fig:fbolcomp}). Although these differences are small, they are statistically significant compared to the random errors ($\lesssim2\%$ in \fbol) for the calibration stars. The higher \fbol{} values result in 1\% higher \teff{} and 4\% higher $L_*$ compared to B12 and corresponding changes in the derived empirical relations between \teff{} and other physical parameters.  We show the relations derived by B12 for \teff-$R_*$ and \teff-$L_*$ next to our own in Figure~\ref{fig:Tfits} to illustrate the size of these differences. Because of other sources of error (e.g, sample size), utilizing B12 values for calibration stars would result in only minor changes in our assigned astrophysical parameters for \kepler targets. When we re-derive the parameters for \kepler planet hosts assuming B12 values for the calibration stars we find that most $R_*$ values change by $<2\sigma$. Nonetheless the changes are systematic and thus should not be ignored. 

We compare PHOENIX model spectra to the visible and NIR spectra of the calibrator stars, and compare the best-fit PHOENIX \teff{} values to the bolometric values. We show that the AGSS version of the PHOENIX models significantly under-predicts the temperatures of M dwarfs, amounting to a difference of 75~K at 3500~K. However, we find that by masking out poorly modeled regions (e.g., at 6500\AA), the CIFIST version of PHOENIX accurately reproduces the bolometrically derived temperatures. With this technique we are able to measure \teff{} accurate to 57~K using moderate-resolution optical spectra. The excellent performance of the CIFIST models highlights how much M dwarf atmospheric models have improved in just the last few years. 

We estimate the parameters of late-type \kepler planet-hosting dwarfs using comparisons of (CIFIST) PHOENIX model spectra to observations, and our calibration between bolometric \teff{} and PHOENIX \teff. We utilize our empirical relations to refine the stars' \teff, $R_*$, $M_*$, and $L_*$. We estimate the radius of each planet using the revised value of $R_*$, and the irradiance it experiences using the revised values of $L_*$ and $M_*$. We determine which of these planets are likely to orbit within the circumstellar HZ. Our spectra of these late-type KOIs are available online.

We find that four planets in or within 1$\sigma$ of the HZ: KOI 854.01, 1298.02, 1686.01, and 2992.01. Only one of these (1686.01) is Earth-size, and KOI 2992.01 has a radius of $\simeq20R_\earth$ and may not be a planet. This suggests that $\eta_{\earth}$ for M dwarfs may need to be revised downward from the \citet{Kopparapu2013b} estimate, although the value depends on the \kepler detection completeness for small planets, which is still being assessed. 

We calculate the irradiation in the 400-700~nm range (useful for oxygenic photosynthesis) incident at the top of the atmosphere for the four HZ planets. Although this is a factor of 3-5 less than that on Earth, it is still adequate to support photosynthetic life. We include spectra of these four stars to aid with modeling of the climate and habitability of these planets.

Our list of HZ stars differs from that of \citet{Gaidos:2013rt}, \citet{Dressing:2013fk} and \citet{Kopparapu2013b}. For example, \citet{Kopparapu2013b} identify KOI 1422.02, 2418.01, and 2626.01 as planets that orbit in their stars' HZ. We find KOI 2626.01 is only 1.2$\sigma$ outside the HZ, and KOI 1422.02 is 4.1$\sigma$ outside the HZ. KOI 2418.01 could be in the HZ depending on the definition used. These discrepancies are largely due to differences in atmospheric models utilized in calculating stellar parameters, and highlight the importance of accurate determinations of \teff{}, $L_*$, and $M_*$ when assessing the habitability of orbiting planets.

Only one of the candidate HZ planets is Earth-size (KOI 1686.01).  However precise radii require not only accurate estimates of $R_*$, but also precise values of the transit depth from \kepler data \citep{Dressing:2013fk} and corrections for limb darkening \citep{Csizmadia:2013}.  Our \teff{} estimates should allow more robust selections of limb darkening laws for future analyses of the \kepler light curves, which in turn can provide better constraints on $R_P/R_*$.

We caution that although our quoted errors for individual \kepler systems are quite small (roughly 57~K in \teff, 7\% in $R_*$, 11\% in $M_*$, and 13\% in $L_*$), unquantified errors may remain. For example, our sample of calibration stars contains only 22 targets, and they are not evenly distributed over \teff. Twenty-two stars is insufficient for a fourth-order (five parameter) fit to the data. Analysis of a much larger sample of stars, such as those with parallaxes and high-resolution spectra \citep[e.g.,][]{Pineda:2013fk}, might shed light on systematic errors present in analyses like ours.

Another issue with our analysis of the \kepler targets is that our calibration sample only covers $-0.4\lesssim \mathrm{[Fe/H]} \lesssim+0.5$. Although none of the stars in our \kepler sample are more metal rich than this, some are as metal poor as [Fe/H]$\simeq$-0.6 \citep{Mann:2013vn, Muirhead:2012zr}. We see no statistically significant effect of metallicity on the \teff-$R_*$ relation, although the existence of subdwarfs, which are smaller than solar-metallicity stars of the same \teff, demonstrates that there must be some dependence of radius on the metallicity at sufficiently low [Fe/H]. It is unclear at what point metallicity becomes an important factor, but this is unlikely to significantly alter our derived stellar parameters for \kepler targets because none of them are subdwarfs. However, we caution against using calibrations for stars significantly more metal poor than [Fe/H]$\simeq$-0.5 where they are untested.

Improvements could be made in our calibrations with additional interferometric measurements for M dwarfs. In particular, the coolest star in our calibration sample is $3238\pm12$~K, but several \kepler planet hosts are near or below that boundary (KOI 2704 \teff{} = 3157$\pm$57, KOI 961 \teff{} = 3241$\pm$57). As mentioned above, none of the stars in the calibration sample have metallicities below [Fe/H]=$-0.40$. However, the number of dwarfs with \teff$<3400$ and/or [Fe/H]$<-0.4$ that are nearby and bright enough for existing facilities like the CHARA array is rather small. Significant improvements in this regime will likely require interferometers with larger apertures or improved instrumentation, e.g. AO for the CHARA array will allow it to reach fainter targets \citep{ten-Brummelaar:2012}.

\acknowledgments
\noindent We thank the anonymous reviewer for their productive comments on the manuscript. We thank Tabetha Boyajian and Kaspar von Braun for their useful discussions on GJ 725B and other stars with interferometrically determined radii, France Allard and Derek Homeier for their advice with the PHOENIX models, and Kimberly Aller for her assistance with some of the aesthetic elements of the manuscript. 

This work was supported by NSF grant AST-0908419, NASA grants NNX10AI90G and NNX11AC33G (Origins of Solar Systems) to EG. SNIFS on the UH 2.2-m telescope is part of the Nearby Supernova Factory project, a scientific collaboration among the Centre de Recherche Astronomique de Lyon, Institut de Physique NuclŽaire de Lyon, Laboratoire de Physique NuclŽaire et des Hautes Energies, Lawrence Berkeley National Laboratory, Yale University, University of Bonn, Max Planck Institute for Astrophysics, Tsinghua Center for Astrophysics, and the Centre de Physique des Particules de Marseille. Based on data from the Infrared Telescope Facility, which is operated by the University of Hawaii under Cooperative Agreement no. NNX-08AE38A with the National Aeronautics and Space Administration, Science Mission Directorate, Planetary Astronomy Program. This paper includes data collected by the \kepler mission. Funding for the \kepler mission is provided by the NASA Science Mission directorate. Some/all of the data presented in this paper were obtained from the Mikulski Archive for Space Telescopes (MAST). STScI is operated by the Association of Universities for Research in Astronomy, Inc., under NASA contract NAS5-26555. Support for MAST for non-HST data is provided by the NASA Office of Space Science via grant NNX09AF08G and by other grants and contracts.

{\it Facilities:} \facility{IRTF}, \facility{UH:2.2m}, \facility{Kepler}

\bibliography{$HOME/dropbox/fullbiblio.bib}

\end{document}